\documentclass[traditarticle]{aa}

\usepackage{graphicx}
\usepackage{stfloats}
\usepackage{txfonts}
\usepackage{xcolor}
\usepackage{booktabs}
\usepackage{natbib}
\bibpunct{(}{)}{;}{a}{}{,}
\usepackage{bm}
\usepackage{hyperref}

\begin{document} 

\title{Gravity’s role in taming the Tayler instability in red giant cores}

\author{Domenico G.~Meduri\inst{1,2} \and
Rainer Arlt \inst{1} \and
Alfio Bonanno \inst{1,2} \and
Giovanni Licciardello \inst{2,3} 
          }

\institute{
Leibniz-Institut f\"{u}r Astrophysik Potsdam (AIP), An der Sternwarte 16, 14482 Potsdam, Germany
\and
INAF -- Osservatorio Astrofisico di Catania, via S.~Sofia 78, 95123 Catania, Italy
\and
Dipartimento di Fisica e Astronomia ``Ettore Majorana'', Universit\`a di Catania, Via S.~Sofia 64, 95123, Catania, Italy
}
\date{}

\abstract{The stability of toroidal magnetic fields in radiative stellar interiors remains a key open problem in astrophysics. We investigate the Tayler instability of purely toroidal fields $B_\phi$ in a nonrotating, stably thermally stratified stellar region in spherical geometry using global linear perturbation analysis and 3D direct numerical simulations.
Both approaches are based on an equilibrium where the Lorentz force is balanced by a gradient of the fluid pressure, and account for gravity and thermal diffusion. The simulations include finite resistivity and viscosity, and cover the full range from stable to highly supercritical regimes for the first time. The linear analysis, global in radius, reveals two classes of unstable nonaxisymmetric modes of azimuthal order $m=1$. High-latitude modes grow at Alfv\'enic rates with short radial length scales, consistent with solutions from the fully local, Wentzel–Kramers–Brillouin (WKB) approximation.
Low-latitude modes, invisible to WKB analyses of a vertical pinch, exhibit larger radial scales and reduced growth rates due to the stabilizing buoyancy. The simulations strongly support these findings and yield threshold toroidal field strengths for both the instability onset and the transition between global and WKB unstable regimes.
These thresholds correspond to the roots of two algebraic equations of the form $B_\phi^{3/4} - a_1\mathcal{A}_1 B_\phi^{1/4}-a_0 \mathcal{A}_0=0$, where $\mathcal{A}_0$ and $\mathcal{A}_1$ are functions of the fundamental fluid properties, and $a_0$ and $a_1$ are coefficients determined from the numerical simulations.
When combined with stellar evolution models of low-mass stars, our results suggest that the outer radiative cores of red giants are generally unstable, while deeper degenerate regions require toroidal fields above $10-100$~kG for instability. These findings provide new constraints for asteroseismic magnetic field detection and angular momentum transport in red giant cores,
and establish a general framework for identifying instability conditions
in other stars with radiative interiors.
}
   \maketitle

\newcommand{\es}{\hat{\vec{e}}_\varpi}
\newcommand{\er}{\hat{\vec{e}}_r}
\newcommand{\etheta}{\hat{\vec{e}}_\theta}
\newcommand{\ephi}{\hat{\vec{e}}_\phi}
\newcommand{\rin}{r_\text{in}}
\newcommand{\rout}{r_\text{out}}
\newcommand{\iu}{\mathrm{i}\mkern1mu}

\section{Introduction}
\label{s:intro}
The origin and evolution of magnetic fields in radiative stellar interiors
remain a fundamental challenge in astrophysics.
 Progress has been hindered by the absence of observational constraints
for any class of stellar objects until recently, when high-precision asteroseismic measurements
from the Kepler mission unveiled the internal magnetic fields
of red giants \citep{Li22}.
The seismic detection relies on mixed modes -- oscillation modes
that act as gravity modes in the core and pressure modes in the envelope of these stars --
which delivered first field strength estimates exceeding 100~kG
due to the suppression of dipolar modes attributed to magnetic fields \citep{Fuller15}.
More recently, asymmetric splittings in the mixed modes frequency spectrum
were measured in 13~low-mass evolved stars, indicating strong radial fields
ranging from 20~to 150~kG \citep{Li22,Li23,Deheuvels23}.
These asteroseismic measurements probe a narrow region of red giant cores
and suggest that the radial field strength decreases from
subgiants to the red giant branch, i.e. with age.
This breakthrough marks a new era in observational
asteroseismology, spurring advances in inverse problems for stellar magnetism \citep[e.g.,][]{Bhattacharya24}
and offering a unique opportunity to test
and refine theoretical models of magnetic field generation and evolution
in low-mass stars.

The high magnetic Reynolds number in radiative stellar interiors
favors the buildup of strong toroidal magnetic fields.
Differential rotation can efficiently wind up even weak poloidal fields
into predominantly toroidal configurations.
Purely toroidal fields in a radiative region are prone to various instabilities, including
the shear-driven magnetorotational instability \citep{Velikhov59,Balbus91}
or the magnetic buoyancy instability \citep{Parker66,Gilman70}.
In strongly stably stratified fluids, it is generally accepted that the dominant instability
is the one caused by electric currents that maintain the toroidal field configuration \citep{Spruit99}.
Current-driven instabilities of pinches and torus configurations were initially studied
for laboratory fusion research, hence neglecting the effect of gravity \citep[e.g.,][]{Freidberg70,Goedbloed72}.
In the astrophysical context, \cite{Tayler73} was the first to consider, in cylindrical geometry, the adiabatic
stability of purely toroidal fields in a stratified, nonrotating star. The author showed that regions
near the poles are always unstable to nonaxisymmetric perturbations with azimuthal wavenumber $m=1$.

The nature of this kink-type instability, known as the Tayler instability,
is significantly altered by the addition of an axial field,
even when its strength is small compared
to the toroidal component \citep{Knobloch92}.
Although it reduces the growth rate, the presence of an axial field introduces
a new resonant instability of the mixed toroidal-poloidal configuration
at high vertical wavenumbers, a behavior confirmed
by numerical simulations \citep{Bonanno08a,Bonanno11}.
The stability of mixed toroidal-poloidal field configurations
in spherical symmetry has been explored through direct numerical simulations \citep{Braithwaite06,Duez10},
with the classic analytical solution by \cite{Prendergast56}
often serving as a reference.
However, recent work has shown that this solution is prone
to resistive instabilities, even under strongly stable stratification,
raising questions about its suitability as a model
for magnetic fields in stars \citep{Kaufman22}.

Stratification and rotation are known to have significant stabilizing effects
on the Tayler instability of purely toroidal fields \citep{Pitts85,Spruit99,Skoutnev24}.
Most linear perturbation analyses adopt a fully local approach
based on the classical Wentzel–Kramers–Brillouin (WKB) approximation
and use a cylindrical geometry valid near the poles of the star.
In contrast, \cite{Bonanno12a} (hereafter BU12)
performed a global analysis in radius,
incorporating both stratification and thermal diffusion in a nonrotating sphere.
Their results showed that while gravity can significantly reduce the instability growth rate $\sigma$,
thermal diffusion prevents complete suppression.
In the stellar regime of highly thermally stable stratification,
where the Brunt-V\"ais\"al\"a frequency $\omega_\text{BV}$ greatly exceeds
the Alfv\'en frequency of the background axisymmetric toroidal field $\omega_\text{A}$, BU12 found that
the growth rates of the least stable eigenmodes
away from the poles -- generally corresponding
to the fundamental eigenmode not captured by fully local analyses -- scale as
\begin{equation}
\label{e:scaling}
\frac{\sigma}{\omega_\text{A}}\sim \left( \frac{\omega_\text{A}}{\omega_\text{BV}}\right)^2 .
\end{equation}
A proper assessment of the dynamical impact of the Tayler
instability in stellar interiors thus requires global analysis
in a spherical geometry,
as demonstrated by BU12 and by \cite{Kitchatinov08},
whose study included disturbances that
are global in latitude.

Although global linear analyses offer valuable insights, they rely
on certain simplifying assumptions and cannot fully replace
direct numerical simulations, which are essential to capture
the full global character of the Tayler instability and
its nonlinear behavior. However, numerical studies in spherical geometry
-- especially under strong stratification -- remain limited.
Early works by \cite{Arlt11} and \cite{Szklarski13} explored the
instability in a spherical shell using the Boussinesq approximation, aiming
to explain the observed surface magnetic fields
of chemically peculiar intermediate-mass stars,
but were restricted to moderately stratified cases.
More recently, \cite{Guerrero19} investigated the impact
of stable stratification on the instability through
simulations in a similar geometry, but within the anelastic
approximation and neglecting viscous and resistive dissipation \citep[for an analysis of the Tayler instability conditions
under the anelastic approximation, see][]{Goldstein19}.
Their simulations confirmed a stabilizing influence
of stratification, although they could not recover the
scaling relation predicted by BU12 (Eq.~\ref{e:scaling}).
Within the same framework, \cite{Monteiro23}
studied the combined effect of rotation and gravity.
In the presence of differential rotation, \cite{Jouve15,Jouve20} and \cite{Meduri24}
investigated numerically the interplay between the magnetorotational instability
of predominantly toroidal fields and the Tayler instability.

In this work, we present direct numerical simulations of the Tayler instability
of a purely toroidal field in a nearly full sphere, extending into the regime
of strong stable stratification.
For the first time, we systematically explore solutions ranging
from subcritical to highly supercritical cases.
Unlike \cite{Guerrero19}, our simulations incorporate both
finite viscosity and magnetic diffusivity, and are initialized with
a physically motivated background toroidal magnetic field
where small deviations from the spherically symmetric equilibrium
are attained by means of a nonspherically symmetric component of the pressure.
Combining global linear perturbation theory and
numerical simulations, we provide
a unified picture of the Tayler instability in a sphere -- capturing
not only the small-scale modes known from WKB analyses,
but also a new class of large-scale unstable fluctuations
that emerge at low latitudes.

The remainder of the paper is organized as follows.
In Sect.~\ref{s:linear_analysis}, we
present a linear stability analysis of a purely toroidal field, global in radius,
generalizing the one of BU12.
We examine the properties of the unstable eigenmodes when varying
the degree of stable stratification,
first neglecting thermal diffusion and then including its effects.
Section~\ref{s:simu} presents 3D direct numerical simulations
in the same spherical geometry and 
initialized with the same background toroidal field as the linear analysis.
We investigate the stability of the relaxed axisymmetric solutions
for different levels of stable stratification and compare
the results with the predictions from the linear analysis.
Our numerical simulation results yield two simple algebraic
relations linking the threshold toroidal magnetic field strengths
for instability onset and the transition between unstable regimes
to key fluid properties.
In Sect.~\ref{s:red_giants}, we combine these numerical
results with stellar evolution models of low-mass stars
to predict corresponding threshold fields in subgiant and red giant cores.
Section~\ref{s:conclusions} closes the paper with a
summary of the results and key conclusions.
\section{Linear perturbation analysis}
\label{s:linear_analysis}
\subsection{Basic equations}
\label{s:lin_eqs}
BU12 considered the stability of a purely axisymmetric toroidal field, $\vec{B}=B_\phi(r,\theta)\,\ephi$,
in a nonrotating stellar radiative region in spherical geometry. They neglected viscous and magnetic diffusion, but included thermal diffusion
with diffusivity $\kappa$.
In this work, $(r,\theta,\phi)$ denote spherical coordinates
with unit vectors $(\er,\etheta,\ephi)$.
The stability analysis is global in radius, but local in latitude and azimuth.

We extend the analysis of BU12 by including the latitudinal variations
of the background toroidal field, which were previously neglected.
In the latitudinal direction, our local approximation requires
$k_\theta \gg H_\theta^{-1}$, where $k_\theta=l/r$, with $l$ the latitudinal wavenumber,
and $H_\theta=r\partial\ln B_\phi/\partial\theta$ is the characteristic
latitudinal scale of the background field.
Stable thermal stratification is considered
under the Boussinesq approximation.
The fluid has a uniform density $\rho$, so gravity varies linearly
with radius, $\vec{g}=-g(r)\er=-g_\text{out} r/r_\text{out}\,\er$.
Here, $g_\text{out}$ is the gravitational acceleration at the outer boundary
of the radiative zone, located at radius $r=r_\text{out}$.

The background toroidal magnetic field varies linearly with the cylindrical radius $\varpi$,
\begin{equation}
\label{e:Bphi_basic}
B_\phi \propto r\sin\theta = \varpi.
\end{equation}
With this field configuration, the local approximation above reads $l\gg\cot\theta$ and therefore the analysis cannot be applied too close to the symmetry axis. This basic state
explicitly satisfies the magnetohydrostatic equilibrium
\begin{equation}
\label{e:basic_state}
    -\frac{\bm{\nabla}P}{\rho} + \bm{g} + \frac{1}{\mu\rho} \left(\bm{\nabla}\times\bm{B}\right)\times\bm{B} = \bm{0},
\end{equation}
which is the relevant force balance in a spherical stellar interior.
Here, $\mu$ is the magnetic permeability of vacuum,
and the total pressure is $P=p(r)+h(r,\theta)$,
where $p(r)$ is the hydrostatic spherically symmetric contribution balancing gravity,
and $h(r,\theta)\propto r^2\sin^2\theta$ is a non-spherically symmetric component whose gradient
balances the Lorentz force.
The field configuration~(\ref{e:Bphi_basic}) satisfies
the classical instability criterion of \cite{Tayler73}
for nonaxisymmetric perturbations with azimuthal wavenumber $m=1$,
\begin{equation}
    \frac{\text{d}\ln B_\phi}{\text{d}\ln \varpi} > -\frac{1}{2},
\end{equation}
which is valid near the symmetry axis of a nonrotating star and in the absence of diffusivities.
We note that WKB analyses typically assume a different background
toroidal field configuration, $B_\phi\propto \sin\theta\cos\theta$,
which can arise from the the shearing of a weak dipolar field by
radial differential rotation
\citep[e.g.,][]{Spruit99,Fuller19,Skoutnev24}.

Following BU12, we consider small amplitude perturbations to the background fluid velocity $\vec{u}=\vec{0}$, magnetic field $\vec{B}$, and temperature $T$, and perform a normal mode analysis, seeking solutions of the form $f^\prime(r)\exp(\sigma t-\iu l \theta - \iu m \phi)$ to the linearized MHD equations. Here, $t$ is time, $\sigma=\text{Re}(\sigma)+\iu\text{Im}(\sigma)$, and $f^\prime(r)$ is a scalar function that stands for one of the components of the perturbed flow velocity $\vec{u}^\prime$, magnetic field $\vec{B}^\prime$, or temperature $T^\prime$. Linearizing the MHD equations around the basic state~(\ref{e:basic_state}) and eliminating all variables in favor of $u_{r}^\prime$ and $T^\prime$ as discussed in BU12, we obtain the two coupled ordinary differential equations
\begin{gather}
    \begin{aligned}
    &\left[\sigma^2 + \omega_\text{A}^2 - \frac{4\omega_\text{A}^4 \cos^2{\theta}}{m^2(\sigma^2 +  \omega_\text{A}^2)}\right] \frac{\text{d}^2 u_{r}^{\prime}}{\text{d}r^2}\\
    &+\frac{4}{r}\left[\sigma^2 + \omega_\text{A}^2 - \frac{4  \omega_\text{A}^4 \cos^2{\theta}}{m^2(\sigma^2 +  \omega_\text{A}^2)} - \iu \frac{l \omega_\text{A}^4 \sin{2\theta}}{m^2(\sigma^2 + \omega_\text{A}^2)} \right] \frac{\text{d} u_{r}^{\prime}}{\text{d}r}\\
    &+\Bigg[ \frac{2}{r^2} \sigma^2 - k_{\perp}^2 (\sigma^2 + \omega_\text{A}^2) + \frac{4\omega_\text{A}^2}{r^2} \Bigg(\frac{k_{\perp}^2}{k_{\phi}^2} -  \frac{k_{\theta}^2}{k_{\phi}^2} \frac{\sigma^2}{\sigma^2 + \omega_\text{A}^2}\\
    &- \frac{2 \omega_\text{A}^2 \cos^2{\theta}}{m^2 (\sigma^2 +  \omega_\text{A}^2)} - \iu \frac{3 l \omega_A^2 \sin{\theta} \cos{\theta}}{m^2 (\sigma^2 +  \omega_\text{A}^2)}\Bigg)\Bigg] u_{r}^{\prime}
    =- k_{\perp}^2 \alpha g \sigma T^{\prime},\label{e:linearv}
    \end{aligned}\\
    \frac{\kappa}{r^2} \frac{\text{d}}{\text{d}r} \left[ r^2 \frac{\text{d} T^\prime }{\text{d}r}\right] - (\sigma + \kappa k_{\perp}^2) T^\prime = \frac{\omega_\text{BV}^2}{\alpha g_\text{out}} u_{r}^\prime.\label{e:linearT}
\end{gather}
In the equations above, the Alfv\'en frequency of the background azimuthal field is
\begin{equation}
    \omega_\text{A}=\frac{B_\phi k_\phi}{(\mu \rho)^{1/2}},
\end{equation}
where $k_\phi=m/r\sin\theta$, and $k_\perp^2=k_\theta^2+k_\phi^2$.
The Brunt-V\"{a}is\"{a}l\"{a} frequency is defined by
\begin{equation}
    \omega_\text{BV}^2=-g_\text{out} \alpha\left(\bm{\nabla}_\text{ad}T-\bm{\nabla} T\right)\cdot\er ,
\end{equation}
where $\alpha$ is the thermal expansion coefficient.
The adiabatic temperature gradient is $\bm{\nabla}_\text{ad}T$ and the actual temperature gradient is subadiabatic, $\bm{\nabla}T > \bm{\nabla}_\text{ad}T$. We employ
the temperature gradient profile
$(\bm{\nabla}_\text{ad}T - \bm{\nabla}T)\cdot \er = A (1/r_\text{out}^{2}- 1/r_\text{in}^{2})^{-1}(1/r_\text{out}^{2} - 1/r^{2})$, which decreases
from $A$ at the inner boundary $r_\text{in}$ to $0$ at the outer boundary $r_\text{out}$.
The perturbed magnetic field $\vec{B}^\prime$ is obtained
from the velocity perturbations through the linearized induction equation:
\begin{equation}
     \frac{\partial \vec{B}^{\prime}}{\partial t} = \bm{\nabla} \times (\vec{u}^{\prime} \times \vec{B}).
\end{equation}

In the limit of large radial wavenumbers, $k_r\gg k_\theta\gg H_\theta^{-1}$, our analysis
is consistent with a purely local approach as expected.
Considering the radial part of the perturbations as $f^{\prime}=\exp{(-\iu k_r r)}$
and retaining only the leading-order terms in $k_r$
in Eqs.~(\ref{e:linearv}) and (\ref{e:linearT}),
we obtain the dispersion relation
\begin{equation}
\label{e:WKB_disp_rel}
\left( \sigma^2 + \omega_\text{A}^2 + \frac{k_{\perp}^2}{k_r^2} \omega_\text{BV}^2 \frac{\sigma}{\sigma + \kappa k_r^2} \right) \left( \sigma^2 + \omega_\text{A}^2 \right) - 4 \frac{\omega_\text{A}^4}{m^2} \cos^2{\theta} = 0 .
\end{equation}
This dispersion relation matches the one obtained by the local analysis of
\citet{Skoutnev24} for our background toroidal field.
As we show below, our linear analysis offers new insight into the influence
of gravity on the Tayler instability in a sphere, overcoming key limitations
of fully local approaches.
\subsection{Numerical results}
\label{s:lin_num}
We nondimensionalize Eqs.~(\ref{e:linearv}) and (\ref{e:linearT})
using the outer boundary radius $r_{\text{out}}$ as reference
length scale. Time is scaled by the Alfv\'en frequency
\begin{equation}
\label{e:lin_def_omeA}
    \omega_{\text{A}0}=\frac{B_0}{(\mu\rho)^{1/2} r_\text{out}} ,
\end{equation}
where $B_0$ is the background toroidal field strength at the equator on the outer boundary.
Temperature perturbations are scaled by the amplitude of the background temperature gradient $A$.
The resulting dimensionless variables are: radius $\tilde{r}=r/r_\text{out}$, radial
velocity perturbation $\tilde{u}_{r}^{\prime}=u_{r}^{\prime}/\omega_{\text{A}0} r_\text{out}$,
and temperature perturbation $\tilde{T}^{\prime}=T^{\prime} / A\, r_\text{out}$.
In this scaling scheme, Eqs.~(\ref{e:linearv}) and (\ref{e:linearT}) read
\begin{gather}
    \begin{aligned}
    &\left(\Gamma^2 + m^2 - \frac{4m^2 \cos^2{\theta}}{\Gamma^2 +  m^2}\right) \frac{\text{d}^2 \tilde{u}_{r}^{\prime}}{\text{d}\tilde{r}^2}\\
    &+ \frac{4}{\tilde{r}}\left(\Gamma^2 + m^2 - \frac{4  m^2 \cos^2{\theta}}{\Gamma^2 +  m^2} - \iu\frac{ l\, m^2 \sin{2\theta}}{\Gamma^2 + m^2}\right) \frac{\text{d} \tilde{u}_{r}^{\prime}}{\text{d}\tilde{r}}\\
    &+\Bigg\{ \frac{2\Gamma^2}{\tilde{r}^2} - \tilde{k}_\perp^2 \left(\Gamma^2 + m^2\right)+ \frac{4m^2}{\tilde{r}^2} \Bigg[\left(\frac{l^2 \sin{\theta}^2}{m^2} + 1 \right)\\
    & -  \frac{l^2 \sin{\theta}^2}{m^2} \frac{\Gamma^2}{\Gamma^2 + m^2} - \frac{2 \cos^2{\theta}}{\Gamma^2 +  m^2}\\
    & - \iu \frac{3 l \sin{\theta} \cos{\theta}}{\Gamma^2 +  m^2}\Bigg]\Bigg\} \tilde{u}_{r}^{\prime}
    = - \delta^2 \tilde{k}_\perp^2 \Gamma \tilde{r}  \tilde{T}^{\prime},\label{e:linearv_nd}
    \end{aligned}\\
    \frac{\epsilon}{\tilde{r}^2}\frac{\text{d} }{\text{d} \tilde{r}} \left(\tilde{r}^2 \frac{\text{d}\tilde{T}^{\prime}}{\text{d} \tilde{r}} \right) - \left(\Gamma + \epsilon \tilde{k}_\perp^2\right) \tilde{T}^{\prime} = \tilde{\omega}_{\text{BV}}^2 \tilde{u}_{r}^{\prime} . \label{e:linearT_nd}
\end{gather}
Here, $\tilde{k}_\perp^2=l^2/\tilde{r}^2 + m^2/\tilde{r}^2 \sin^2{\theta}$, $\Gamma=\sigma/\omega_{\text{A}0}$, and the dimensionless
Brunt-V\"{a}is\"{a}l\"{a} frequency is $\tilde{\omega}_{\text{BV}}=\omega_\text{BV}/\omega_{\text{BV}0}$, where
we defined $\omega_{\text{BV}0}^2= \alpha g_\text{out} A$.
In the equations above, there are two dimensionless parameters
quantifying the relative importance of the three key timescales
of the problem. The first is the stratification parameter
\begin{equation}
    \delta=\frac{\omega_{\text{BV}0}}{\omega_{\text{A}0}},
\end{equation}
and the second is the ratio of the thermal diffusion rate $\omega_T=\kappa/r_\text{out}^2$ to the Alfv\'en frequency,
\begin{equation}
\epsilon=\frac{\omega_T}{\omega_{\text{A}0}}.
\end{equation}
In radiative stellar interiors, the Brunt-V\"{a}is\"{a}l\"{a} frequency
is typically the highest of the three frequencies,
reflecting strong stable stratification.
The thermal diffusivity $\kappa$ is usually much larger than both the kinematic
viscosity and magnetic diffusivity, justifying our initial assumption of an inviscid and perfectly electrically conducting fluid.
Provided that the toroidal magnetic field is not very weak, we therefore
expect the following ordering of timescales
\begin{equation}
    \omega_{\text{BV}0} \gg \omega_{\text{A}0} \gg \omega_{T}
\end{equation}
or, in terms of dimensionless parameters,
\begin{equation}
    \epsilon \ll 1 \ll \delta.
\end{equation}
To avoid singularities at the center, we introduce a small inner core of radius $\tilde{r}_\text{in}=r_\text{in}/r_\text{out}=0.1$.
We employ impenetrable flow boundary conditions, that is $\tilde{u}_{r}^\prime=0$
at both $\tilde{r}_\text{in}=0.1$ and $\tilde{r}_\text{out}=1$.
The temperature perturbations $\tilde{T}^\prime$ are set to zero at both boundaries.

Hereafter we fix the latitudinal wavenumber to $l=10$
and $\epsilon$ to $10^{-2}$.
For this choice of $l$, the local approximation adopted in our analysis
is not strictly valid near the poles, specifically for colatitudes $\theta\lesssim 5^\circ$.
However, we find no significant difference in the results
when comparing $\theta = 5^\circ$ with a higher value of
$10^\circ$, where the local approximation is well satisfied.
Equations~(\ref{e:linearv_nd}) and (\ref{e:linearT_nd}),
together with the boundary conditions above,
define a nonlinear eigenvalue problem. We solve this system numerically using
an eight-order centered finite-difference scheme for the radial derivatives.
A radial grid of 4000~points is employed, which we have verified
to provide consistent numerical results across the full range of
stable stratification values considered.

Instability occurs when $\text{Re}(\Gamma)$, the real part of $\Gamma$, is positive.
For simplicity of notation, we define $\text{Re}(\Gamma)=\Gamma$ and omit
the superscript~$\,\tilde{}\,$ for nondimensional quantities henceforth.
Note that the coefficients of the linear equations depend on the
colatitude $\theta$, implying that the growth rate also varies
with this quantity, i.e., $\Gamma=\Gamma(\theta)$.
We confirmed that instability arises for the azimuthal wavenumber $m=1$,
while both the axisymmetric mode ($m=0$) and higher-order modes ($m>1$)
remain stable, as expected for the adopted background field configuration.
The next two sections examine the behavior of this $m=1$ instability
under increasing stable stratification, by varying $\delta$ from 0 to $10^3$.
We first consider the case without thermal diffusion
and then include its effects.
\subsubsection{Diffusionless case}
\label{s:lin_notherm}
\begin{figure}
\includegraphics[width=0.95\linewidth]{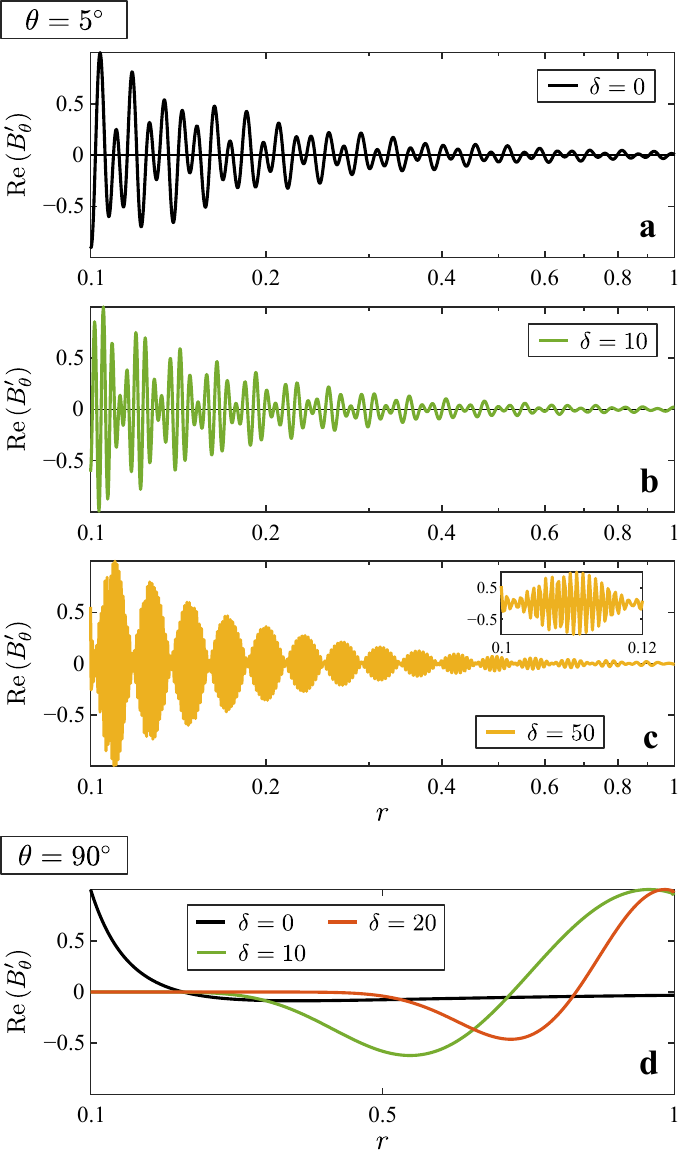}
\centering
\caption{Radial profiles of the real parts of
selected unstable eigenmodes $B_{\theta}^\prime$
in the absence of thermal diffusion ($\epsilon=0$).
Panels (a)-(c) show the high-latitude eigenmodes ($\theta=5^\circ$)
with the nearly adiabatic growth rate of $\Gamma=0.95$
for different values of the stratification parameter $\delta$ (see legend insets).
The inset in (c) provides a zoom on the eigenmode
between $r=0.10$ and $0.12$. Panel (d) displays the most
unstable eigenmodes at the equator ($\theta=90^\circ$);
no unstable modes are found for $\delta\geq 25$. The horizontal
axis is logarithmic in (a)-(c) and linear in (d).
}
\label{f:lin_noeps}
\end{figure}
\begin{figure}[t]
\includegraphics[width=\hsize]{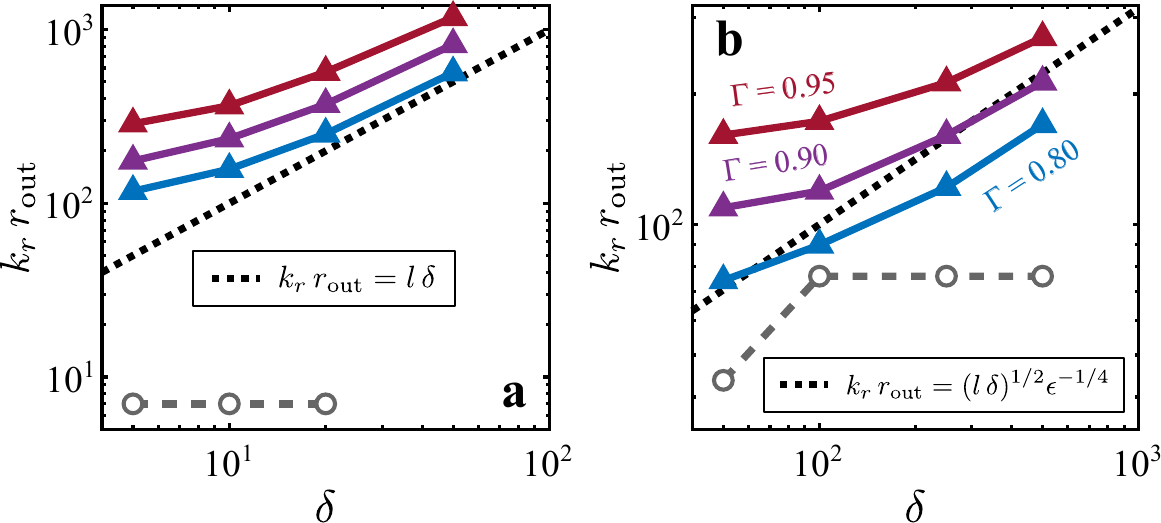}
\centering
\caption{Radial wavenumber $k_r$ of unstable eigenmodes,
evaluated from the real part of $B_\theta^\prime$,
as a function of $\delta$ for (a) no thermal diffusion ($\epsilon=0$)
and (b) finite thermal diffusion ($\epsilon=10^{-2}$).
Triangles connected by solid lines indicate eigenmodes
at colatitude $\theta=5^\circ$ with growth rates $\Gamma=0.8$ (blue),
0.9 (purple), and 0.95 (red).
Selected eigenmodes of the $\Gamma=0.95$ tracks in (a) and (b)
are displayed in Fig.~\ref{f:lin_noeps}a-c and Fig.~\ref{f:lin_eps}a-c,
respectively.
The dotted lines denote the minimum unstable wavenumber
predicted by the local linear theory (Eqs.~\ref{e:kr_WKB_eps0} and \ref{e:kr_WKB}).
Open gray points connected by dashed lines correspond to the equatorial eigenmodes
(Fig.~\ref{f:lin_noeps}d and dashed lines in Fig.~\ref{f:lin_eps}d).
}
\label{f:lin_kr}
\end{figure}
We begin by considering the case without thermal diffusion, $\epsilon=0$.
We found two classes of unstable eigenmodes, the first operating at high latitudes
and the second one in the equatorial region. The first class shares properties 
expected for the modes known from the WKB approximation, while the second one is
completely distinct from the first, as we shall demonstrate below.

We now describe the first class of eigenmodes, focusing on
a colatitude of $\theta=5^\circ$.
Figure~\ref{f:lin_noeps}a-c presents the radial profiles
of the real part of the eigenmodes $B_{\theta}^\prime$
with a nearly adiabatic growth rate of $\Gamma=0.95$,
for different values of the stratification parameter $\delta$.
For all values of $\delta$ explored, the eigenmodes are
largely confined in the inner half of the radial domain.
This spatial structure is characteristic of all eigenmodes
across the growth rate spectrum.
The imaginary parts of the eigenmodes closely resemble the real ones
and are therefore not shown here.

Stable stratification impedes radial displacements and
thereby determines the characteristic instability length scale
in that direction.
Classical local linear analyses show that,
in the diffusionless limit,
the radial wavenumber $k_r$ of the $m=1$ Tayler modes growing at the adiabatic
rate $\omega_\text{A}$ satisfy \citep{Spruit99,Skoutnev24}
\begin{equation}
\label{e:kr_WKB_eps0_dim}
k_r > k_\theta \frac{\omega_\text{BV}}{\omega_\text{A}} .
\end{equation}
For the background state considered here,
$\omega_\text{BV}/\omega_\text{A}\sim 1/r^2$,
implying that $k_r$ decreases with increasing radius.
This leads to a radial variation of the eigenmode scale,
with longer wavelengths in the outer regions where stratification
is weaker -- consistent with the
morphology observed in Fig.~\ref{f:lin_noeps}a-c.

Using the definitions of the previous section,
Eq.~(\ref{e:kr_WKB_eps0_dim}) can be estimated as
\begin{equation}
\label{e:kr_WKB_eps0}
k_r \,r_\text{out} > l\,\delta,
\end{equation}
where $l=10$.
We estimated the characteristic radial wavenumber of
each eigenmode as $k_r=\pi/\lambda_r$, where $\lambda_r=D/n$
is the mean half-wavelength. Here, $n$ denotes
the number of radial nodes (zero crossings) within the domain
of extent $D=r_\text{out}-r_\text{in}$.
Figure~\ref{f:lin_kr}a presents $k_r$ as a function of $\delta$
for the eigenmodes $\text{Re}(B_{\theta}^\prime)$ with growth rates
$\Gamma=0.8$, $0.9$, and $0.95$.
For all these high growth rates, which are close to the adiabatic value,
the radial wavenumbers exceed the WKB limit given by Eq.~\ref{e:kr_WKB_eps0} (dotted line)
and converge towards the expected scaling for $\delta >10$.
This confirms that our eigenvalue problem, in the limit of large $k_r$,
recovers the WKB dispersion relation (Eq.~\ref{e:WKB_disp_rel}),
thereby validating the numerical solver.

When moving away from the symmetry axis towards lower latitudes,
the amplitude of the component of the buoyancy force
in the cylindrical radial direction $\es$ -- which counteracts
the destabilizing Lorentz force of the background field --
progressively increases, thereby stabilizing
the WKB-like eigenmodes discussed above.
Such a geometrical effect on the stabilizing role of gravity
in a sphere was already noted by \cite{Goossens80}
in their diffusionless stability analysis.
For $\theta \gtrsim 70^\circ$, we find no unstable WKB-like
eigenmode when $\delta\geq 20$.
At the equator ($\theta=90^\circ$), gravity completely
opposes the unstable cylindrical radial motions,
and we find no instability at all for $\delta \geq 25$.
For $\delta<25$, stable stratification is weaker, and instability
exists, but in the form of a new class of eigenmodes, as anticipated above.
The most unstable magnetic field fluctuations are large scaled, exhibiting only
one or a few radial nodes (Fig.~\ref{f:lin_noeps}d).
As $\delta$ increases, the most unstable eigenmode
shifts outward, toward regions where stable stratification is
locally weaker.
Unlike the WKB-like modes at high latitudes, which maintain
growth rates near $\omega_{\text{A}0}$ for all
values of stratification explored,
the most unstable equatorial eigenmodes show a sharp decline
in growth rate, from $0.98$ at $\delta=0$
to $0.19$ at $\delta=20$.
\begin{figure}[t]
\includegraphics[width=0.95\linewidth]{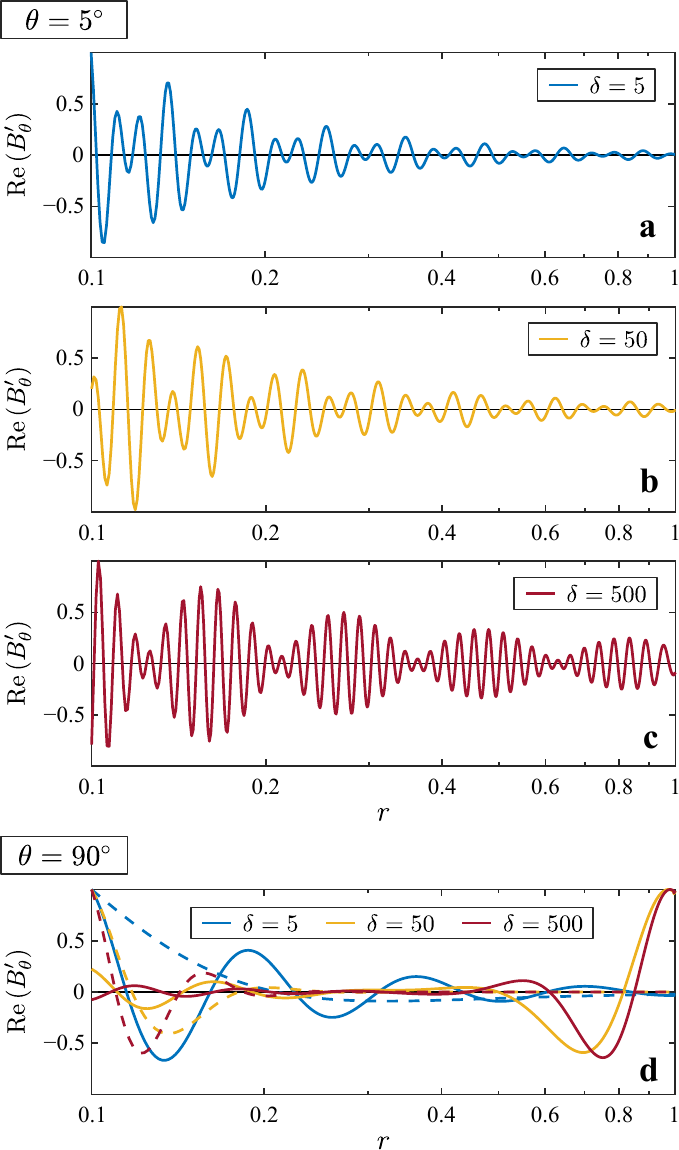}
\centering
\caption{Same as Fig.~\ref{f:lin_noeps} but for $\epsilon=10^{-2}$.
In panel~(d), dashed and solid lines correspond to the localized (type~I)
and distributed (type~II) equatorial eigenmodes, respectively (see text for details).
Type~I eigenmodes are the most unstable (see Fig.~\ref{f:lin_grwr_delta}
for their growth rate values).
The horizontal axis is logarithmic in all panels.
}
\label{f:lin_eps}
\end{figure}
\subsubsection{Effect of thermal diffusivity}
\label{s:lin_therm}
We now examine the influence of thermal diffusion on the
instability, considering $\epsilon=10^{-2}$.
Thermal diffusion weakens the effect of the stabilizing
buoyancy force by damping the amplitude of the temperature
perturbations in the buoyancy force on the right-hand side of
Eq.~(\ref{e:linearv_nd}), thereby modifying both the radial structure
and growth rates of the unstable eigenmodes.
As in the diffusionless case discussed above, we find
two distinct classes of unstable eigenmodes: rapidly growing, WKB-like modes
at high latitudes, and slower, large-scale eigenmodes near the equator.

In the presence of thermal diffusion, unstable WKB modes growing
at the adiabatic rate $\omega_\text{A}$ have radial wavenumbers
that satisfy the condition \citep{Spruit99,Skoutnev24}
\begin{equation}
    \label{e:kr_WKB_dim}
    k_r^4 > k_\theta^2 \frac{\omega_\text{BV}^2}{\omega_\text{A} \kappa} .
\end{equation}
Using the definitions introduced earlier, this condition can be recast
in nondimensional form as
\begin{equation}
    \label{e:kr_WKB}
    k_r r_\text{out} > (l\,\delta)^{1/2}\epsilon^{-1/4}.
\end{equation}
We first demonstrate that, at high latitudes, the less stable eigenmodes
from our linear analysis share salient characteristics
with such pure WKB modes.
As in the diffusionless case, eigenmodes at colatitude $\theta=5^\circ$
and for $\delta\leq 50$ display large growth rates and remain
confined to
the inner half of the radial domain. This is illustrated
in Fig.~\ref{f:lin_eps}a,b, which shows the radial profiles
of $\text{Re}(B_\theta^\prime)$ with nearly adiabatic
growth rates $\Gamma=0.95$.
Due to the effect of thermal diffusion, the radial length scales of these
eigenmodes are larger than those of the corresponding diffusionless case,
for all values of $\delta$ explored -- as expected
(cf.~Fig.~\ref{f:lin_eps}b and Fig.~\ref{f:lin_noeps}c for the case at $\delta=50$).
The radial variation of the eigenmode length scales, which arises
from the background stratification profile, appears less pronounced
than in the absence of diffusion.
For $\delta>50$, the dominant eigenmodes exhibit a different morphology,
with significant amplitudes extending into the outer
regions of the radial domain (Fig.~\ref{f:lin_eps}c for $\delta=500$).
The radial wavenumber $k_r$, estimated from $\text{Re}(B_\theta^\prime)$
as discussed in the previous section, falls above
the predicted WKB limit (Eq.~\ref{e:kr_WKB}; dotted line in Fig.~\ref{f:lin_kr}b)
for the eigenmodes with nearly
adiabatic growth rates $\Gamma = 0.95$, as expected (Fig.~\ref{f:lin_kr}b, red triangles).
The WKB scaling, $k_r r_\text{out}\propto\delta^{1/2}$,
is closely approached for $\delta > 200$.

Due to the geometrical reasons discussed above, the
stabilizing effect of stratification on the unstable motions
increases with colatitude and peaks at the equator.
As in the diffusionless case, no WKB-like eigenmodes are found
at colatitudes $\theta\gtrsim 70^\circ$. However, the instability persists in the form
of a distinct class of eigenmodes characterized by large radial length scales
and significantly reduced growth rates.
At the equator ($\theta=90^\circ$), these global eigenmodes exhibit
two distinct types of morphologies.
The first, which we refer to as type~I, corresponds to the
most unstable modes and is characterized by confinement near
the inner boundary (Fig.~\ref{f:lin_eps}d, dashed lines).
The radial wavenumber $k_r$ of these eigenmodes increases with
stratification at first, but then saturates beyond $\delta=10^2$
(Fig.~\ref{f:lin_kr}b, circles; Fig.~\ref{f:lin_eps}d, dashed lines).
Their growth rate strongly decreases
with stratification, from $\Gamma=0.98$ for $\delta=5$
to a value as small as $0.012$ in the highly stratified case at $\delta=500$.
The second type of equatorial eigenmodes, referred to as type~II hereafter,
have relatively lower growth rates
than type~I modes (for example, $\Gamma=0.41$ and $6\times 10^{-3}$
for $\delta=50$ and 500, respectively). Similar to the eigenmodes
reported by BU12, type~II eigenmodes are distributed primarily
in the outer half of the radial domain at higher levels of stable stratification
(Fig.~\ref{f:lin_eps}d, solid lines).

\begin{figure}[t]
\includegraphics[width=0.8\linewidth]{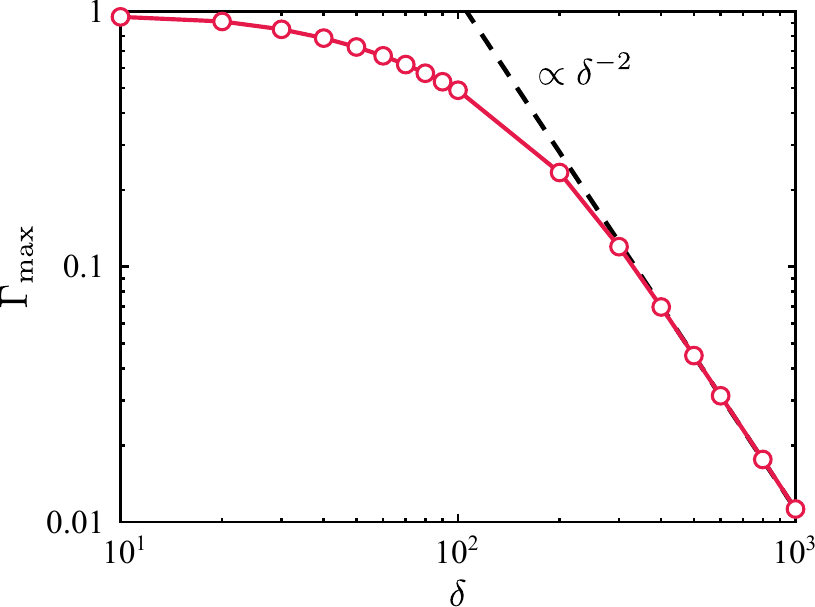}
\centering
\caption{Dimensionless growth rates of the
type~I (most unstable) equatorial eigenmodes, $\Gamma_\text{max}$,
as a function of the stratification parameter $\delta$ for $\epsilon=10^{-2}$.
The dashed line indicates the scaling
$\Gamma_\text{max}\propto \delta^{-2}$ predicted by \citet{Bonanno12a}.}
\label{f:lin_grwr_delta}
\end{figure}
Figure~\ref{f:lin_grwr_delta} presents the growth rate
of the type~I equatorial eigenmodes,
$\Gamma_\text{max}$, as a function of the stratification parameter $\delta$.
For weak up to moderate stratification values of $\delta=200$,
the growth rate slowly decreases as stratification
increases, but always remaining of the order of the background Alfv\'en frequency
$\omega_{\text{A}0}$.
In the highly stratified regime where $\delta>200$, the growth rate
decreases more steeply, following the scaling
\begin{equation}
    \Gamma_\text{max}\propto \delta^{-2}
\end{equation}
already predicted by BU12 (dashed line).
The growth rates of the type~II equatorial eigenmodes
show the same scaling with $\delta$ and are therefore not presented here.

Unlike fully local approaches, our linear analysis 
successfully captures the global radial structure of the instability
and allowed us to identify a new class of large scaled unstable modes
developing at low latitudes.
To extend the analysis beyond the local approximation in latitude
and to include viscous and resistive effects,
we now turn to self-consistent
direct numerical simulations, which we present in the next section.
\section{Direct numerical simulations}
\label{s:simu}
\subsection{Governing equations}
As in the linear analysis above, we consider a MHD fluid confined
to a wide-gap spherical shell of aspect ratio
$\chi=\rin/\rout=0.1$, where $\rin$
and $\rout$ denote the inner and outer boundary radii, respectively.
The fluid has uniform density $\rho$, gravity varies linearly with radius,
$\vec{g}=-g(r)\er=-g_\text{out}r/r_\text{out}\er$,
and stable stratification is considered under the Boussinesq approximation
as in the linear analysis.
Here, $g_\text{out}$ is the gravitational acceleration at the outer boundary.
Unlike the linear analysis, we consider a viscous fluid with
finite electrical conductivity. The kinematic viscosity, magnetic diffusivity,
and thermal diffusivity of the fluid are $\nu$, $\eta$, and $\kappa$, respectively.

We employ the mid-shell radius $r_0=\rin+D/2$
as reference length scale and the magnetic diffusion time $r_0^2/\eta$
as reference timescale. Here, $D=\rout-\rin$ denotes the spherical shell gap.
The magnetic field is scaled in units of $(\mu\rho)^{1/2}\eta/r_0$, hence
the dimensionless magnetic field coincides with the Lundquist number
\begin{equation}
    \text{Lu}=\frac{B^\ast r_0}{(\mu\rho)^{1/2}\eta}.
    \label{e:def_Lu}
\end{equation}
The fluid velocity is expressed in units
of $\eta/r_0$. The scale for the temperature
is $\Delta T= T_\text{out}-T_\text{in}>0$,
the imposed background temperature contrast between the isothermal outer and
inner boundaries that establishes stable stratification.
The non-hydrostatic pressure $\Pi$ is scaled with $\rho\eta^2/r_0^2$.
In this scaling scheme, the dimensionless equations governing the evolution of
the fluid velocity $\vec{u}$, the magnetic field $\vec{B}$, and the temperature
perturbations $\vartheta$ around the background temperature $T_\text{c}$ are:
\begin{gather}
    \begin{aligned}
    \frac{\partial\vec{u}}{\partial t} + (\vec{u}\cdot\bm{\nabla})\vec{u} = &-\bm{\nabla}\Pi
    +\delta_0^2\,\text{Lu}_0^2\frac{r}{\rout}\vartheta\,\er + \text{Pm}\bm{\nabla}^2\vec{u}\\
    &+\left(\bm{\nabla}\times\vec{B}\right)\times\vec{B},\label{eq:NS}
    \end{aligned}
    \\
    \frac{\partial\vec{B}}{\partial t} =\bm{\nabla}\times (\vec{u}\times \vec{B}) + \bm{\nabla}^2\vec{B},\label{eq:ind}\\
    \frac{\partial\vartheta}{\partial t} + (\vec{u}\cdot\bm{\nabla})\vartheta + (\vec{u}\cdot\bm{\nabla}) T_\text{c} = \frac{\text{Pm}}{\text{Pr}}\bm{\nabla}^2\vartheta \label{eq:temp}.
\end{gather}
The flow and the magnetic field obey solenoidality,
\begin{equation}
\bm{\nabla}\cdot\vec{u}=0\quad\text{and}\quad \bm{\nabla}\cdot\vec{B}=0.
\end{equation}
The background temperature $T_\text{c}$ is the solution to the
(dimensionless) heat conduction equation $\bm{\nabla}^2 T_\text{c}=0$:
\begin{equation}
    T_\text{c}(r)=-c_1/r+c_2,
\end{equation}
where the coefficients $c_1=2\chi/(1-\chi^2)$ and $c_2=1+(1-\chi)^{-1}$
are determined by the imposed inner and outer boundary values of
$T_\text{in}/\Delta T=1$ and $T_\text{out}/\Delta T=2$, respectively.
The Brunt-V\"{a}is\"{a}l\"{a} frequency $N$ is related to this temperature profile by
\begin{equation}
    N^2(r)=g\alpha\frac{\text{d}T_\text{c}}{\text{d}r},
\end{equation}
where $\alpha$ is the coefficient of thermal expansion. We define
a reference value for the Brunt-V\"{a}is\"{a}l\"{a} frequency, $N_0$, as
\begin{equation}
    N_0^2=g_{\text{out}}\alpha\frac{\Delta T}{r_0}.
\end{equation}
At both spherical boundaries, $r=\rin$ and $r=\rout$, we
impose impenetrable, stress-free boundary conditions
for the flow, and perfect conductor conditions for the magnetic field.

In Eqs.~(\ref{eq:NS})--(\ref{eq:temp}), there
are four dimensionless control parameters. The first is
the stratification parameter
\begin{equation}
   \delta_0 = \frac{N_0}{\omega_{\text{A}0}},
\end{equation}
where $\omega_{\text{A}0}=B_0^\ast/(\mu\rho)^{1/2}r_0$ is the Alfv\'en frequency
based on the initial magnetic field strength $B_0^\ast$, which will
be defined below. Note that this definition of $\omega_{\text{A}0}$
differs from Eq.~(\ref{e:lin_def_omeA}), which is used throughout
Sect.~\ref{s:linear_analysis}.
The second parameter is the initial Lundquist number
\begin{equation}
    \text{Lu}_0=\frac{B^\ast_0\, r_0}{(\mu\rho)^{1/2}\eta} .
\end{equation}
The remaining two parameters are the Prandtl number, $\text{Pr}=\nu/\kappa$,
and the magnetic Prandtl number, $\text{Pm}=\nu/\eta$.

To solve the problem above, we employ the open-source pseudo-spectral MHD code
MagIC\footnote{https://magic-sph.github.io} \citep{Wicht02,Schaeffer13}.
The numerical method is described in detail
in \cite{Christensen07} and we therefore mention only the essentials here.
The code uses a poloidal-toroidal decomposition of the vector fields
$\bm{u}$ and $\bm{B}$. The associated poloidal and toroidal potentials, along with
the temperature, are expanded using spherical harmonic functions
in the latitudinal and azimuthal directions.
In the radial direction, a Chebyshev collocation method with $N_r$
grid points is employed.
The discretized equations are evolved in time using an implicit-explicit scheme,
with the nonlinear terms handled explicitly.
The numerical spatial resolutions are set to adequately capture
the characteristic radial and angular length scales of the instability
during both its linear growth and nonlinear evolution (see Sect.~\ref{s:nonaxi}).
For the highest degrees of stratification explored, we employ a maximum
resolution of $N_r=321$ radial grid points and $\ell_\text{max}=m_\text{max}=213$,
where $\ell_\text{max}$ and $m_\text{max}$ denote the maximum spherical
harmonic degree and order, respectively.
\begin{table}
\caption{\label{t:sim_runs}Input parameters
and output diagnostics of the numerical simulation runs
(see main text for definitions).}
\centering
\begin{tabular}{@{}lccccccr@{}}
\toprule\toprule
Name & \multicolumn{1}{c}{$\delta_0$} & $\text{Lu}_0$ & $\text{Pr}\cdot 10^2$ & $t_1\,\omega_{\text{A}0}$ & 
\multicolumn{1}{c}{$\overline{\delta}_1$} & $\overline{\text{Lu}}_1$ & $\sigma/\overline{\omega}_{\text{A}1}$\\ \midrule
TU0$^{(1)}$   & 0   & 917   & -                   & 26.6   & 0   & 677  & 1.09\\
TU1           & 0   & 611   & -                   & 50.6   & 0   & 406  & 1.09\\
TU2           & 0   & 367   & -                   & 31.5   & 0   & 242  & 1.06\\
TU3           & 0   & 244   & -                   & 51.4   & 0   & 122  & 0.90\\
TU4           & 0   & 122   & -                   & 61.6   & 0   & 30   & 0.28\\
TU5           & 0   & 61    & -                   & 40.9   & 0   & 10   & stable\\
TU6           & 0   & 31    & -                   & 26.1   & 0   & 3    & stable\\
TS1$^{(1)}$   & 8   & 917   & $1$           & 24.3   & 3.5 & 681  & 0.94\\
TS2$^{(1)}$   & 63  & 917   & $1$           & 24.2   & 27  & 682  & 0.77\\
TS3$^{(1)}$   & 126 & 917   & $1$           & 26.0   & 54  & 680  & 0.60\\
TS4$^{(1)}$   & 209 & 917   & $1$           & 25.2   & 90  & 681  & 0.43\\
TS5$^{(1)}$   & 300 & 917   & $1$           & 26.9   & 130 & 678  & 0.28\\
TS6$^{(1,2)}$ & 419 & 917   & $1$            & 24.5   & 180 & 682  & 0.16\\
TS7$^{(1)}$   & 628 & 917   & $1$             & 27.7   & 272 & 678  & 0.07\\
TS8$^{(1)}$   & 900 & 917   & $1$             & 27.3   & 321 & 678  & 0.04\\
TS9$^{(2)}$   & 419 & 458   & $2$    & 11.6   & 180 & 342  & stable\\
TS10$^{(2)}$  & 419 & 1833  & $0.5$    & 46.0   & 180 & 1367 & 0.47\\
TS11$^{(2)}$  & 419 & 3667  & $0.25$  & 91.2   & 180 & 2735 & 0.68\\
TS12$^{(2)}$  & 419 & 7333  & $0.125$ & 173.2  & 180 & 5481 & 0.85\\
TS13          & 628 & 458   & $2$    & 13.8   & 272 & 339  & stable\\
TS14          & 739 & 306   & $3$    & 9.3    & 321 & 226  & stable\\
TS15          & 739 & 397   & $2.31$ & 12.1   & 321 & 293  & stable\\
TS16          & 739 & 611   & $1.5$  & 18.3   & 321 & 452  & stable\\
TS17          & 63  & 57    & $16$  & 1.7    & 27  & 42   & stable\\
TS18          & 63  & 458   & $2$    & 12.7   & 27  & 340  & 0.55\\
\bottomrule
\end{tabular}
\tablefoot{Superscripts $^{(1)}$ and $^{(2)}$ in the run names
indicate a run pertaining to the simulation subsets 1 and 2, respectively
(see Fig.~\ref{f:regime_diagr}).
The time at which the nonaxisymmetric perturbations
are introduced is $t_1$.
The final column reports the growth rate of the $m=1$ mode, with
``stable'' indicating decaying perturbations.
All runs have a magnetic Prandtl number of $\text{Pm}=1$.
For all stratified runs, the ratio of the thermal diffusion
rate to the mean Alfv\'en frequency at the perturbation
time is $\overline{\epsilon}_1 = 0.15$.}
\end{table}
\subsection{Initial magnetic field configuration and values of dimensionless parameters}
As initial condition for the magnetic field, we consider a purely axisymmetric azimuthal field
of the form
\begin{equation}
    \bm{B}(t=0)=\text{Lu}_0\,r\sin\theta\,\ephi .
\end{equation}
Before studying the stability of such a field configuration,
we first investigate its temporal evolution
when varying the strength of the stable stratification
and the field amplitude.
The stratification parameter, $\delta_0$, spans values
from 0 (corresponding to an unstratified flow) up to 900,
while the Lundquist number, $\mathrm{Lu}_0$, ranges from subcritical values,
for which no instability is observed, to approximately 7300.
The magnetic Prandtl number $\text{Pm}$ is fixed to $1$.
When varying the Lundquist number $\text{Lu}_0$ at fixed stratification $\delta_0$,
to maintain a constant ratio of the thermal diffusion rate
to the Alfv\'en frequency at $\epsilon_0 = \omega_T / \omega_{\text{A}0} = 0.11$,
we adjust the Prandtl number $\text{Pr}$ accordingly.
Here, the thermal diffusion rate is defined as $\omega_T = \kappa / r_0^2$,
and $\epsilon_0$ can be expressed
in terms of the basic input parameters as $\epsilon_0=\text{Pm}\big/\text{Pr}\,\text{Lu}_0$.
The input parameters of all simulation runs are listed in
Table~\ref{t:sim_runs}.

In contrast to the outer regions of red giant cores, the Prandtl number and
the magnetic Prandtl number attain relatively high values
in the inner, electron degenerate layers -- values that are accessible
to numerical simulations.
Stellar evolution models indicate typical values of $\text{Pr} \sim 10^{-3}$ \citep{Garaud15}
and $10^{-1} \lesssim \text{Pm} \lesssim 10$ \citep{Ruediger15}
for the degenerate cores of red giant branch stars with masses near that of the Sun.
The value $\text{Pm}=1$ adopted in this study lies within this range,
while the employed values of $\text{Pr}$ are, in most cases, only one order
of magnitude larger (see Table~\ref{t:sim_runs}).

Hereafter, we denote the azimuthal and meridional averages
of an arbitrary function $f(r,\theta,\phi,t)$ with angular brackets and
an overbar respectively,
\begin{gather}
    \langle f \rangle=\frac{1}{2\pi}\int_0^{2\pi}f\,\text{d}\phi\quad\text{and}\\
    \overline{f} = \frac{1}{\text{A}}\int_0^\pi\int_{r_{\text{in}}}^{r_{\text{out}}}f\,r\,\text{d}r\,\text{d}\theta ,
\end{gather}
where $\text{A}$ is the area of half the meridional plane.
The nonaxisymmetric components of the flow velocity
and magnetic field are denoted by a prime symbol. For example, the nonaxisymmetric
azimuthal magnetic field is $B^\prime_\phi=B_\phi-\langle B_\phi \rangle$.
\subsection{Background axisymmetric configuration}
\label{s:axisym}
\begin{figure}[t]
\centering
\includegraphics[width=0.85\hsize]{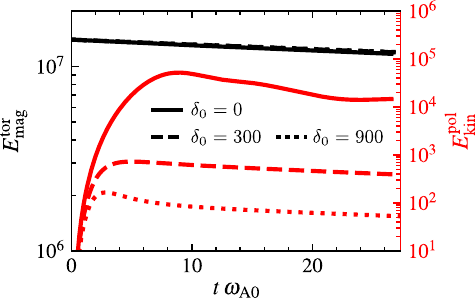}
\caption{Temporal evolution of the volume-averaged
toroidal magnetic energy (black lines) and poloidal kinetic energy (red lines).
Results are shown for the unstratified run ($\delta_0=0$)
and two stratified runs ($\delta_0=300$ and $900$).
The Lundquist number is $\text{Lu}_0=917$. Both the flow and
magnetic field remain purely axisymmetric throughout the evolution.
}
\label{f:axi_ener}
\end{figure}
\begin{figure}[b]
\centering
\includegraphics[width=0.9\hsize]{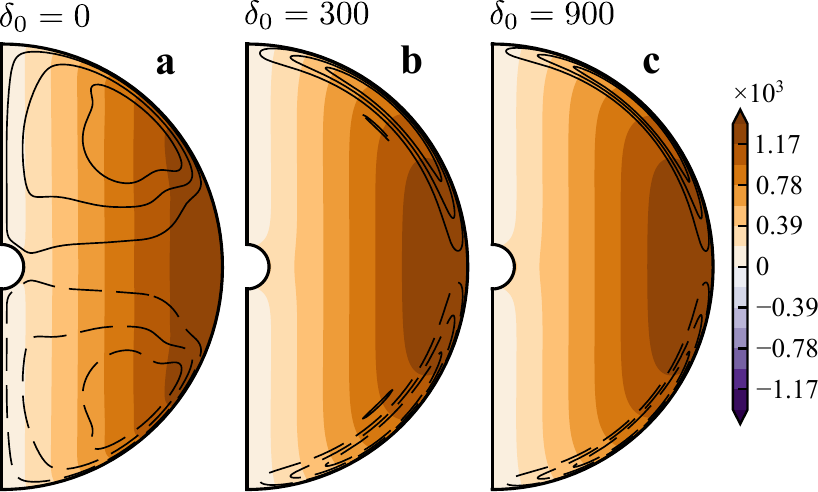}
\caption{Snapshots of the three axisymmetric simulation runs presented
in Fig.~\ref{f:axi_ener}. Snapshot times $t\omega_{\text{A}0}$ are 26.6
for (a) $\delta_0=0$, 26.9 for (b) $\delta_0=300$, and 27.3 for (c) $\delta_0=900$.
Color contours show the
axisymmetric azimuthal field $\langle B_\phi\rangle$. Black contour lines
represent the meridional circulation, with solid and dashed lines
indicating clockwise and counterclockwise flows, respectively.
}
\label{f:axi_bphi_snapsh}
\end{figure}
\begin{figure*}
\centering
\includegraphics[width=0.80\hsize]{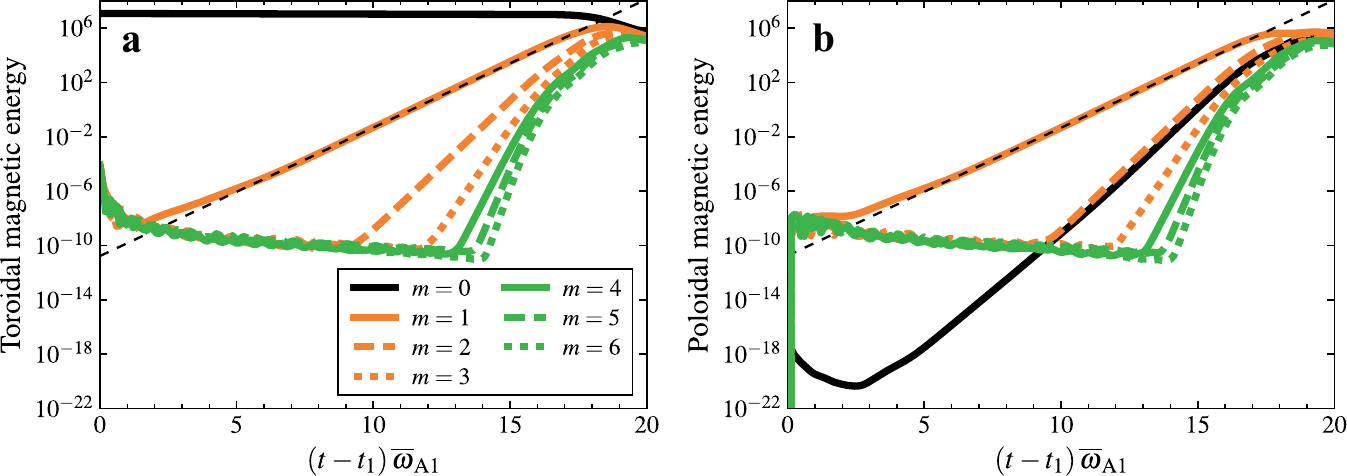}
\caption{Temporal evolution of the (a) toroidal and (b) poloidal
magnetic energies for the azimuthal modes $m=0$ to $m=6$
in the unstratified run TU0.
The black dashed lines show exponential fits to the energy evolution of
the $m=1$ mode
within the time interval $6.5\leq (t-t_1)\,\overline{\omega}_{\text{A}1}\leq 10.5$.
These fits are used to determine the instability growth rates $\sigma$ discussed in the text.
}
\label{f:unstrat_ener_m}
\end{figure*}
We first explore the slow, long-term evolution of the
axisymmetric toroidal field due to resistive effects,
by focusing on a Lundquist number of $\text{Lu}_0 = 917$
and varying the strength of stable stratification.

The temporal evolution of the volume integrated toroidal magnetic energy,
$E_\text{mag}^\text{tor}=\frac{1}{2}\int B_{\text{tor}}^2\,\text{d}V$,
is presented by the black lines in Fig.~\ref{f:axi_ener}
for the unstratified case ($\delta_0=0$),
an intermediately stable stratification ($\delta_0=300$),
and a highly stable stratification ($\delta_0=900$).
The magnetic field remains axisymmetric throughout its evolution
and is purely toroidal, as no poloidal component
is initialized in our simulations.
The toroidal field decays
exponentially due to Ohmic diffusion as expected.
The snapshots in Fig.~\ref{f:axi_bphi_snapsh} demonstrate
that the axisymmetric azimuthal field, $\langle B_\phi \rangle$,
preserves the cylindrical symmetry of its initial state.
Its configuration is modified exclusively towards the spherical
boundaries to conform to the perfect conductor boundary conditions.

We observe an initial rapid growth of the poloidal kinetic energy
$E_\text{kin}^\text{pol}=\frac{1}{2}\int u_{\text{pol}}^2\,\text{d}V$
(red lines in Fig.~\ref{f:axi_ener}) driven by the development of
meridional flows induced by the Lorentz force.
These flows subsequently relax to their end-state configuration within some
Alfv\'en travel times.
The amplitude of these flows decreases with increasing $\delta_0$,
consistent with the restoring buoyancy force inhibiting
radial motions.
In the unstratified run, the meridional flow exhibits
a single large-scale circulation cell in each hemisphere (black isocontour lines
in Fig.~\ref{f:axi_bphi_snapsh}a). With increasing stable stratification,
the meridional flow becomes confined near the outer boundary, forming
thin circulation cells that locally advect the azimuthal field (Fig.~\ref{f:axi_bphi_snapsh}b,c).
Over longer timescales, the meridional flow amplitude decays,
mirroring the behavior of the toroidal field
and confirming their magnetic origin.
\subsection{Stability to nonaxisymmetric perturbations}
\label{s:nonaxi}
We investigate the stability of the background axisymmetric
configurations in response to weak nonaxisymmetric perturbations.
The perturbations are introduced at times $t=t_1$
(indicated in Table~\ref{t:sim_runs}) and consist
of spatially uncorrelated white noise in the nonaxisymmetric ($m>0$)
component of the toroidal magnetic field potential.
The root mean square (RMS) amplitude of these perturbations
is at least 4 orders of magnitude lower than that
of the background axisymmetric azimuthal field.

As demonstrated in the previous section, the axisymmetric
azimuthal field strength decreases over time, and its morphology shows
mild variations with $\delta_0$.
To account for these changes, we define a local version of the
stratification parameter at the perturbation time $t_1$,
\begin{equation}
    \delta_1 = \frac{N}{\omega_\text{A1}},
\end{equation}
where $\omega_{\text{A}1}=B_{\phi 1}^\ast/(\mu\rho)^{1/2}\,r_0$
is the Alfv\'en frequency of the axisymmetric azimuthal field
at the perturbation time
$B_{\phi 1}^\ast(r,\theta)=\langle B_{\phi}^\ast(r,\theta,\phi,t=t_1)\rangle$.
The parameter $\delta_1$ exhibits approximate cylindrical symmetry,
with some modulations introduced
by the chosen Brunt-V\"{a}is\"{a}l\"{a} frequency profile,
and varies roughly as $1/r^2\sin\theta$ within the fluid domain.
To characterize the degree of stable stratification at the perturbation time,
we define a mean value of the stratification parameter as
$\overline{\delta}_1=\overline{N}\big/\overline{\omega}_{\text{A}1}$,
which is reported in Table~\ref{t:sim_runs} for all runs.
Similarly, the characteristic azimuthal field strength at the perturbation
time is represented by the mean Lundquist number
$\overline{\text{Lu}}_1=\overline{B}^\ast_{\phi 1}\,r_0/(\mu\rho)^{1/2}\eta$.
This parameter can be interpreted as the ratio of the mean Alfv\'en frequency
$\overline{\omega}_{\text{A}1}$
to the magnetic diffusion rate $\omega_\eta=\eta/r_0^2$.

We analyze the stability of a set of axisymmetric simulation runs in
which stable stratification increases up to $\overline{\delta}_1 = 321$,
while the dimensionless field strength spans from $\overline{\text{Lu}}_1 = 3$ to 5481.
The ratio of the thermal diffusion rate to the mean Alfv\'en frequency
at the perturbation time is $\overline{\epsilon}_1=\omega_T/\overline{\omega}_{\text{A}1}$.
This parameter is related to the other dimensionless quantities
by $\overline{\epsilon}_1=\text{Pm}/\text{Pr}\, \overline{\text{Lu}}_1$
and its value is of 0.15 in all runs explored.
In the following section, we focus on a representative case without stratification,
where the Tayler instability exhibits its known WKB-like behavior.
The influence of stable stratification and the various instability
regimes identified are presented in Sect.~\ref{s:simu_strat}.
\subsubsection{Tayler instability with no stratification}
\label{s:simu_unstrat}
We begin by examining the stability of the unstratified run
described in Sect.~\ref{s:axisym}.
The nonaxisymmetric perturbations are introduced
at time $t_1=26.6\,\omega_{\text{A}0}^{-1}$
when $\overline{\text{Lu}}_1=677$ (see Fig.~\ref{f:axi_bphi_snapsh}a
for the background field configuration).
This value of $\overline{\text{Lu}}_1$ exceeds the threshold for
instability onset by more than a factor of 20 (see Table~\ref{t:sim_runs}).
This simulation is hereafter designated as run TU0
and establishes a reference case
for the characterization of the $m=1$ Tayler instability.

\begin{figure*}[b]
\centering
\includegraphics[width=\hsize]{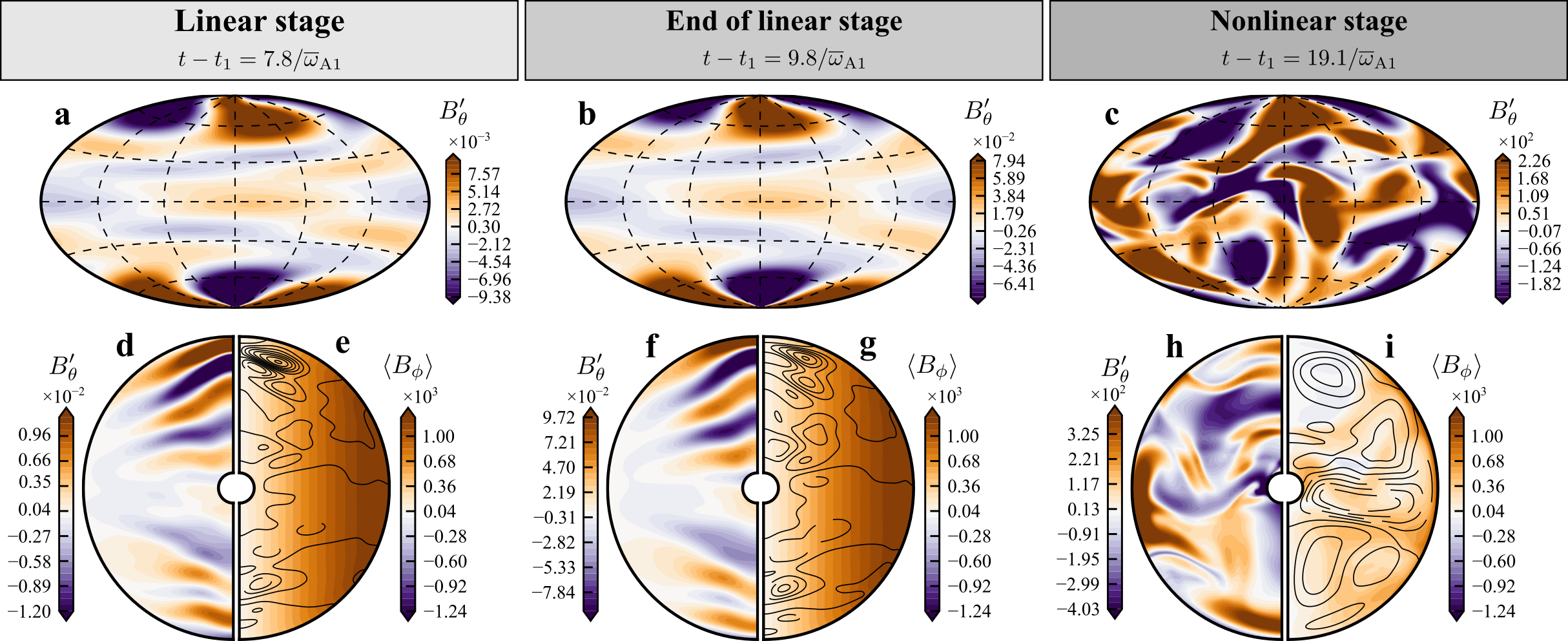}
\caption{Snapshots of the magnetic field fluctuations
and axisymmetric field for the unstratified run TU0.
The three leftmost, middle, and rightmost panels show the
linear stage of the instability, the end of the linear stage,
and the nonlinear stage, respectively. (a-c)~Hammer projections
of the nonaxisymmetric latitudinal field $B_\theta^\prime$
at radius $r/\rout=0.7$. (d,f,h) Meridional slices
of $B_\theta^\prime$ at longitude $\phi=90^\circ$.
(e,g,i) Color isocontours depict the axisymmetric azimuthal field,
$\langle B_\phi\rangle$, with overlaid black contour lines
indicating the axisymmetric poloidal field.
}
\label{f:unstrat_snapsh}
\end{figure*}
Figure~\ref{f:unstrat_ener_m} shows the temporal evolution of the toroidal and poloidal
magnetic energies for the azimuthal modes $m=0$ to $m=6$ in this run.
Approximately $2/\overline{\omega}_{\text{A}1}$ after the perturbation time, the first
mode to grow exponentially is $m=1$, as expected for the Tayler instability.
The linear growth rate of the RMS toroidal field associated to this mode
is $\sigma_{m=1}=1.1\,\overline{\omega}_{\text{A}1}$, when evaluated in the period
$(t-t_1)\,\overline{\omega}_{\text{A}1}=6.5-10.5$ (dashed black line in Fig.~\ref{f:unstrat_ener_m}a).
This value closely matches the rate calculated from the evolution of the
$m=1$ poloidal field (solid orange line in Fig.~\ref{f:unstrat_ener_m}b).
The background axisymmetric ($m=0$) azimuthal field decays slowly due to Ohmic diffusion,
with its evolution appearing nearly stationary during the linear
growth phase of the instability (solid black line in Fig.~\ref{f:unstrat_ener_m}a).

As expected, snapshots of the nonaxisymmetric
latitudinal field $B_\theta^\prime$
during the linear phase of the instability growth reveal
a distinct $m=1$ azimuthal symmetry (Fig.~\ref{f:unstrat_snapsh}a,b).
All the other nonaxisymmetric field and flow components exhibit
the same symmetry and are therefore not shown here.
There is no azimuthal drift of the nonaxisymmetric flow and field
components, as expected in the absence of rotation \cite[e.g.,][]{Ruediger12}.
The instability remains stationary also in the vertical direction
and predominantly localizes
in a wedge-shaped region around the vertical symmetry axis (Fig.~\ref{f:unstrat_snapsh}d,f).
This latter characteristic aligns with predictions from our linear analysis,
which indicates that the growth rates of the unstable eigenmodes
decrease with the colatitude $\theta$, and is also consistent
with previous results obtained from local linear stability studies in spherical geometry \citep{Goossens80}.
Our linear analysis also indicates that the characteristic
radial wavenumber of the instability increases from the equator towards
higher latitudes (black lines in Fig.~\ref{f:lin_noeps}a,d),
which qualitatively matches the structure of the instability in run TU0.
We refrain from making a direct quantitative comparison between
the numerical simulation results and the linear analysis
due to the limitations inherent in the latter's approximations.
In particular, the linear analysis may become questionable
at higher latitudes near the symmetry axis, where the local approximation
in the latitudinal direction is only partially satisfied.
Additionally, the linear analysis neglects viscous and
resistive effects, which would certainly influence the spectrum
of the unstable radial modes if they were included.

Finally, we note that the structure of the instability in run TU0
closely resembles that observed in global numerical studies of the Tayler
instability in cylindrical geometry, which utilize the same background azimuthal
field configuration as employed here \citep{Ruediger12,Ji23}.

Figure~\ref{f:unstrat_ener_m} shows that modes $m>1$, as well as the
axisymmetric poloidal field (black line in the right panel),
are subcritical initially. They begin to grow 
only at times $t-t_1 \gtrsim 10/\overline{\omega}_{\text{A}1}$, as
a result of nonlinear energy transfer from the large-amplitude $m=1$ mode.
A simple weakly nonlinear analysis of the ideal induction equation
by \cite{Ji23} predicts that the nonlinear growth rates of the
$m=0$ and $m=2$ modes are $2\,\sigma_{m=1}$.
Higher chains of nonlinear interaction provide a nonlinear growth
rate of $m\,\sigma_{m=1}$ for modes $m > 2$.
Figure~\ref{f:unstrat_ener_m}b demonstrates that the nonlinear growth
of the $m=0$ and $m=2$ poloidal field occurs at the same rate of $2.2\,\overline{\omega}_{\text{A}1}$,
which is twice the rate of the $m=1$ mode, as anticipated.
The nonlinear growth rates of the higher order modes
in Fig.~\ref{f:unstrat_ener_m}
are also in good agreement with the weakly nonlinear theory predictions.
For simplicity, we hereafter denote the growth rate of the $m=1$
mode as $\sigma$.

The instability shows an apparent saturation
at $t-t_1\approx 18/\overline{\omega}_{\text{A}1}$
when the nonaxisymmetric toroidal energy
reaches values comparable to those of the background azimuthal field (Fig.~\ref{f:unstrat_ener_m}a),
with the nonaxisymmetric poloidal and toroidal energies roughly in equipartition.
The field solution looses its simple azimuthal symmetry and transitions into a
turbulent state, characterized by small-scale structures (Fig.~\ref{f:unstrat_snapsh}c,h).
In this nonlinear stage, the axisymmetric field relaxes
into a mixed poloidal-toroidal configuration (Fig.~\ref{f:unstrat_snapsh}i), which no longer exhibits instability
and slowly decays by resistive effects, along with the instability fluctuations.
A detailed characterization of the nonlinear behavior of the instability
is beyond the scope of this work and will be addressed in a future study.
In the following section, we focus on the effect of stable stratification
on the linear properties of the instability.
\subsubsection{Effect of stable stratification and instability regimes}
\label{s:simu_strat}
As discussed in Sect.~\ref{s:linear_analysis}, stable stratification has a strong impact on the Tayler instability. In the presence of thermal diffusion, local linear theory predicts that Tayler modes growing at an adiabatic rate $\sigma\approx \omega_\text{A}=B_\phi^*\big/(\mu\rho)^{1/2} r$ have radial wavenumbers satisfying Eq.~\ref{e:kr_WKB_dim}:
\begin{equation}
    k_r^4 > \frac{N^2}{\omega_\text{A}\kappa\, r^2},
\end{equation}
when using the notation of this section
and assuming $k_\theta\approx 1/r$. The minimum radial wavenumber allowed by the effective stratification response is then
\begin{equation}
\label{e:kN}
    k_N = \left( \frac{N^2}{\omega_\text{A}\kappa\,r^2}\right)^{1/4}.
\end{equation}
At larger wavenumbers the instability is damped when resistive or viscous
effects become dominant, that is
\begin{equation}
    \eta k_r^2 > \sigma .
\end{equation}
This relation is valid when $\nu\leq \eta$ (or $\text{Pm}\leq 1$).
For $\text{Pm}>1$, the magnetic diffusivity $\eta$ in the above relation
can be substituted with the kinematic viscosity $\nu$, given that the local dispersion relation
is symmetric with respect to $\nu$ and $\eta$ \citep{Skoutnev24}.
Considering $\sigma\approx \omega_\text{A}$, the maximum unstable radial wavenumber is thus
\begin{equation}
\label{e:keta}
    k_\eta = (\omega_\text{A}/\eta)^{1/2} .
\end{equation}
Instability occurs when $k_N < k_\eta$.

\begin{figure}
\centering
\includegraphics[width=\hsize]{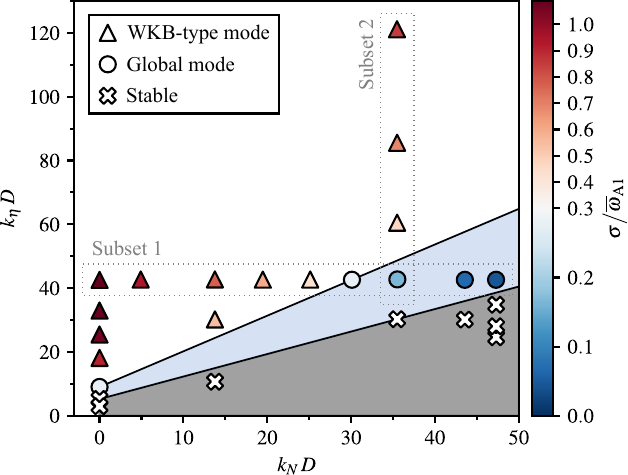}
\caption{Instability diagram showing the
resistive wavenumber $k_\eta$ as a function
of the stratification wavenumber $k_N$, both scaled by the shell gap $D$.
Individual simulation runs are represented by symbols: crosses
indicate decaying nonaxisymmetric perturbations, while circles and
triangles mark unstable global and WKB-type $m=1$ modes, respectively.
The symbol color codes the instability growth rate $\sigma$.
The gray and blue shaded regions denote the stable regime and
the global mode regime, respectively, based on the simulation data.
The boundaries of these shaded regions are defined by
Eq.~\ref{e:sim_stab_line} (stability line) and Eq.~\ref{e:sim_glob_line}
(transition line between the global and WKB regimes).
Thin dotted horizontal and vertical rectangles highlight runs
in subset~1 ($\overline{\text{Lu}}_1\approx 680$)
and subset~2 ($\overline{\delta}_1=180$).
}
\label{f:regime_diagr}
\end{figure}
By adopting $r=r_0$ as the typical radial location of the instability
in our simulations
and $\omega_\text{A}=\overline{\omega}_{\text{A}1}$
as the characteristic Alfv\'en frequency of the background toroidal field,
Eq.~(\ref{e:kN}) and (\ref{e:keta}) can be rewritten, scaled with the shell gap $D$, as
\begin{equation}
\label{e:kN_nondim}
    k_N D = \overline{\delta}_1^{1/2} \overline{\text{Lu}}_1^{1/4} (\text{Pr}/\text{Pm})^{1/4} D/r_0
\end{equation}
and
\begin{equation}
\label{e:keta_nondim}
    k_\eta D = \overline{\text{Lu}}^{1/2}_1\,D/r_0 .
\end{equation}
Fig.~\ref{f:regime_diagr} presents $k_\eta D$ as a function of
$k_N D$ for all our simulation runs.
Crosses denote runs where the nonaxisymmetric perturbations decay,
from which we derive the empirical stability line
\begin{equation}
\label{e:sim_stab_line}
    k_\eta D = 5.22 + 0.71\, k_N D .
\end{equation}
This equation defines the boundary of the gray shaded
region in the figure.
It is important to note that this relationship, with its distinct
offset and slope, deviates from the marginal stability line
$k_\eta = k_N$ predicted by the local linear theory.

Triangles and circles in Fig.~\ref{f:regime_diagr} represent
simulation runs where the perturbations exhibit exponential growth.
Specifically, triangles denote unstable fluctuations resembling
WKB modes, while circles correspond to the global equatorial
modes identified in our linear analysis, as we will discuss further.
In all these runs, the $m=1$ azimuthal mode is the first to
grow exponentially after introducing the perturbations, similar
to the reference unstratified case TU0 (Sect.~\ref{s:simu_unstrat}).
Higher azimuthal modes display analogous dynamical evolutions
to those observed in run TU0 and are thus not discussed here.
Far from the instability threshold, where $k_\eta \gg k_N$ and
resistive effects are negligible, the growth rates of the $m=1$ mode,
indicated by the symbol color in Fig.~\ref{f:regime_diagr},
are consistent with $\sigma\approx\overline{\omega}_{\text{A}1}$,
which aligns with the WKB prediction.
Throughout this section, $\sigma$ is calculated from the temporal evolution
of the volume-averaged toroidal magnetic energy associated to the $m=1$ mode, as described in Sect.~\ref{s:simu_unstrat}.

Due to their lower growth rates, global modes primarily become apparent in
the simulations when approaching the stable regime, that
is when resistive effects begin to damp the small-scale WKB modes.
The transition line between the WKB and global mode regimes, derived from
the simulation data, is
\begin{equation}
\label{e:sim_glob_line}
    k_\eta D = 9.01 + 1.12\, k_N D ,
\end{equation}
which defines the upper boundary of the blue shaded region
in Fig.~\ref{f:regime_diagr}.

\begin{figure}
\centering
\includegraphics[width=\hsize]{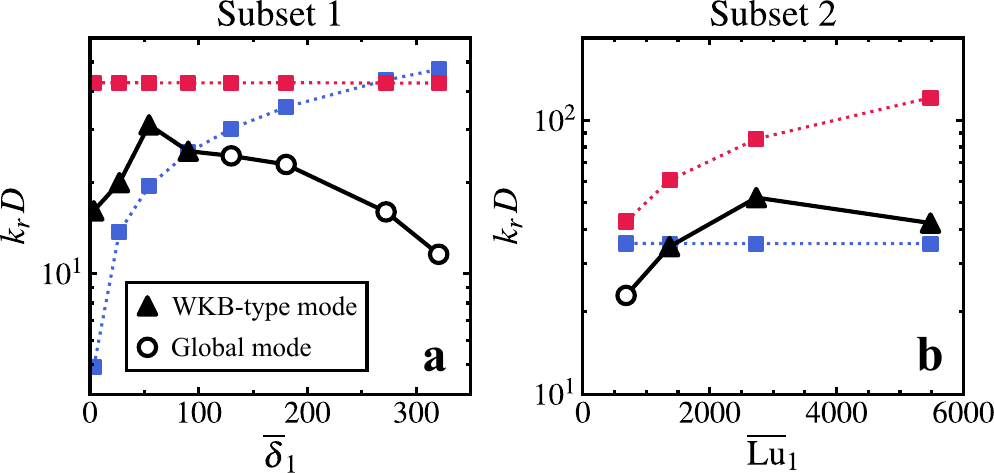}
\caption{Comparison of the radial wavenumbers $k_r$
observed in the numerical simulation runs of (a) subset~1 and (b) subset~2
(triangles and circles; see also Fig.~\ref{f:regime_diagr}) with the WKB predictions.
Blue and red squares show, respectively, the minimum ($k_N$)
and maximum ($k_\eta$) unstable wavenumbers as
predicted by the WKB theory (Eqs.~\ref{e:kN_nondim} and \ref{e:keta_nondim}).
}
\label{f:kr}
\end{figure}
\begin{figure*}[b]
\centering
\includegraphics[width=\hsize]{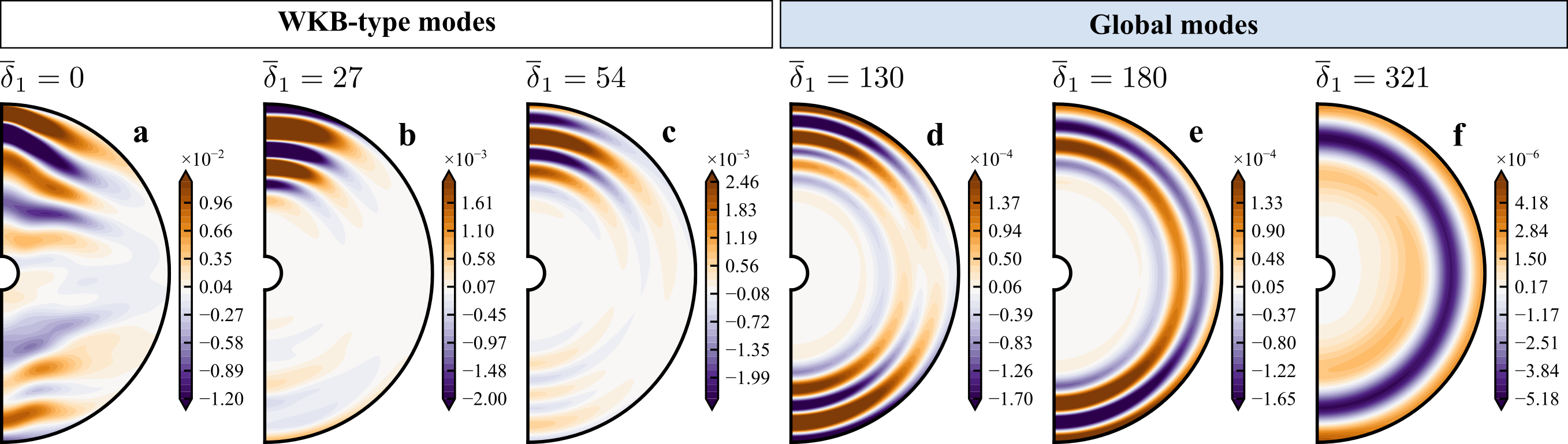}
\caption{Meridional slices of the nonaxisymmetric
latitudinal magnetic field, $B_\theta^\prime$, during
the linear growth phase of the instability
for six simulation runs of subset~1 in Fig.~\ref{f:regime_diagr}, spanning different values
of the stratification parameter $\overline{\delta}_1$.
WKB-type modes are identified for $\overline{\delta}_1\leq 54$, while
global modes emerge for $\overline{\delta}_1\geq 130$.
The leftmost panel corresponds to the unstratified run TU0 described in Sect.~\ref{s:simu_unstrat}.
}
\label{f:snapsh_Strat}
\end{figure*}
We now explore how the properties of the Tayler modes
identified in the simulations change from the WKB
to the global mode regime by following
two distinct paths in the parameter space.
The first path fixes the resistive wavenumber
at $k_\eta\approx 42/D$ (or $\overline{\text{Lu}}_1\approx 680$)
and gradually increases $k_N$ by varying
the stratification parameter $\overline{\delta}_1$.
The simulation runs for this first path, hereafter referred to as subset~1, are enclosed
by the thin horizontal rectangle in Fig.~\ref{f:regime_diagr}.
The second path towards the global mode regime
is defined by a fixed $k_N = 35/D$ (or $\overline{\delta}_1=180$) and
a decreasing $k_\eta$, achieved by varying the Lundquist number
$\overline{\text{Lu}}_1$. Runs belonging to this second path are
designated as subset~2 (thin vertical rectangle in Fig.~\ref{f:regime_diagr}).

For subset~1 runs with $k_N \leq 25/D$ (or $\overline{\delta}_1 \leq 90$),
the instability growth rate is close to the adiabatic one
with $\sigma \gtrsim 0.4\,\overline{\omega}_{\text{A}1}$ (triangles in the horizontal
rectangle of Fig.~\ref{f:regime_diagr}; see also Table~\ref{t:sim_runs}).
The magnetic field fluctuations are localized from high to intermediate latitudes,
in a wedge-shaped region around the vertical symmetry axis,
as expected for WKB modes (Fig.~\ref{f:snapsh_Strat}a-c).
We estimated the radial wavenumbers of the instability in subset~1 runs
from snapshots of $B_\theta^\prime$ during the linear growth phase.
Specifically, they are calculated as $k_r=\pi/\lambda_r$, where $\lambda_r$
is the characteristic half-wavelength. We estimated $\lambda_r$ as the mode 
of the distribution of the lengths between two successive nodes in the 
radial profiles of $B_\theta^\prime$ obtained at a colatitude $\theta=30^\circ$ 
and for all longitudes. The wavenumbers observed in these runs
consistently fall within the theoretical WKB limits ($k_N < k_r < k_\eta$) and, as expected,
increase with stratification, as demonstrated by Fig.~\ref{f:kr}a (triangles).

As $\overline{\delta}_1$ increases to $130$,
where $k_\eta D$ is only about $1.3\, k_N D$, resistive effects
begin to dampen the WKB modes. We identify a new class
of unstable $m=1$ fluctuations which, unlike the previous WKB-type modes,
are nearly constant in latitude and start also to be active in the
equatorial region (Fig.~\ref{f:snapsh_Strat}d).
The growth rate of these modes also reduces to $0.28\,\overline{\omega}_{\text{A}1}$
and their radial wavenumber is $k_r\approx 25/D$, which
falls below the minimum wavenumber $k_N\approx 30/D$
predicted by the local linear theory (Fig.~\ref{f:kr}a).
Further increasing $\overline{\delta}_1$ leads to fluctuations
with consistently lower growth rates (circles in subset~1 runs of Fig.~\ref{f:regime_diagr})
and increasingly larger radial length scales, which
further deviate from the theoretical WKB predictions
(Fig.~\ref{f:kr}a, circles; Fig.~\ref{f:snapsh_Strat}e,f).
These findings align with our linear analysis, as the global
equatorial eigenmodes also display large radial
length scales and growth rates that diminish with stable stratification (Sect.~\ref{s:lin_therm}).

Simulation runs in subset~2 at $\overline{\delta}_1=180$
(vertical rectangle in Fig.~\ref{f:regime_diagr}) exhibit analogous behavior
when moving towards the global mode regime by decreasing the Lundquist number.
At the largest $\overline{\text{Lu}}_1$ of 5481 explored (or $k_\eta = 121/D$) the instability
growth rate is near adiabatic, with $\sigma=0.85\,\overline{\omega}_{\text{A}1}$.
The magnetic fluctuations are concentrated at high latitudes (not shown)
and are small-scaled, with a characteristic radial
wavenumber of $k_r=45/D$, falling within the expected WKB limits
(Fig.~\ref{f:kr}b).
As $\overline{\text{Lu}}_1$ decreases, $k_r$
progressively declines within the range of unstable WKB wavenumbers,
until global modes emerge at the lowest $\overline{\text{Lu}}_1$ of $680$
(circle in Fig.~\ref{f:kr}b).

\begin{figure}
\centering
\includegraphics[width=\hsize]{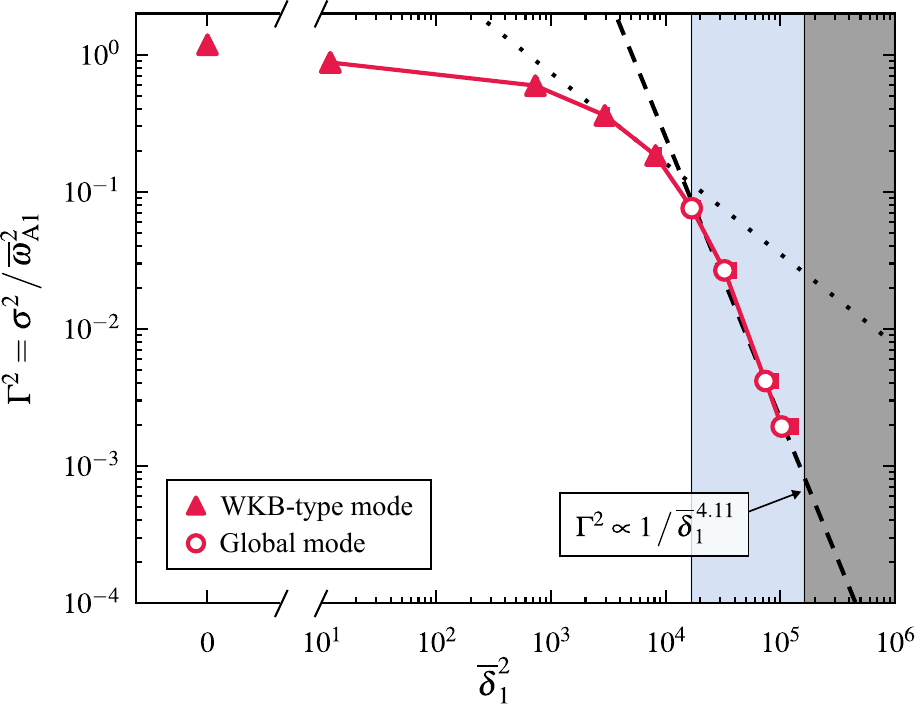}
\caption{Square of the normalized instability growth rate, $\Gamma^2$, as a function
of $\overline{\delta}_1^2$ for the simulation runs of subset~1
(see Fig.~\ref{f:regime_diagr}).
The dashed line shows the best-fit power law for runs
in the global mode regime ($\overline{\delta}_1^2>10^4$; circles):
$\Gamma^2\propto 1/\overline{\delta}_1^{4.1}$.
The dotted line indicates the best-fit for runs
in the moderately stratified regime with WKB-type modes
(triangles in $10^3 <\overline{\delta}_1^2 < 10^4$): $\Gamma^2\propto 1/\overline{\delta}_1^{1.3}$.
The gray and blue shaded areas mark, respectively, the stable and
global mode regimes as identified in Fig.~\ref{f:regime_diagr}.
The thick horizontal bars associated with each data point
indicate the range of $\overline{\delta}_1$
values attained during the linear phase of the instability growth (see text for details).
For the unstratified case and the first five stratified runs, these bars
are smaller than the symbol size.
}
\label{f:growth_rate}
\end{figure}
Finally, we detail the variation of the instability growth rate
with stratification, focusing on the simulation runs in subset~1.
Figure~\ref{f:growth_rate} presents the square of the normalized growth rate, $\Gamma^2=\sigma^2/\overline{\omega}_{\text{A}1}^2$,
as a function of $\overline{\delta}_1^2$ for these runs.
For $\overline{\delta}_1^2 \lesssim 10^3$, where stratification is weak,
the growth rate $\Gamma$ only slightly decreases from the adiabatic value of 1
observed in the unstratified run TU0.
In this regime, as previously demonstrated, we clearly
identified WKB-like modes characterized by large radial wavenumbers (Fig.~\ref{f:kr}a).
At intermediate values of the stratification parameter,
$10^3 < \overline{\delta}_1^2 < 10^4$, we find stronger variations
in the growth rate, yet still identify WKB-like modes.
A power law fit $\Gamma^2=c\,\overline{\delta}_1^{2b}$
to the simulation data points in this
stratification range yields $c=71$ and an exponent $b=-0.66$.
This fit is shown as a dotted line in Fig.~\ref{f:growth_rate}.
The value of $b$ is close to the one found in the
numerical simulations of \cite{Guerrero19}, which is $-0.95$, suggesting
that their study has focused on WKB-type Tayler modes.

For $\overline{\delta}_1^2 > 10^4$,
where global modes are identified,
the best fitting parameters are $c=4.07\times 10^7$ and $b=-2.05$
(dashed line in Fig.~\ref{f:growth_rate}).
This value of $b$ is in excellent agreement with
the prediction by BU12, which is also confirmed
by our linear analysis in Sect.~\ref{s:lin_therm}.
In this regime, the instability grows at such a slow rate that the amplitude
of the background axisymmetric field $\langle B_\phi \rangle$ decreases
during the linear evolution of the instability, thereby leading to an increase
in $\overline{\delta}_1$ over time.
The thick horizontal bars associated to the data points in Fig.~\ref{f:growth_rate}
indicate the range of values of $\overline{\delta}_1$
attained during the entire linear phase of the instability, and demonstrate
that such variations do not impact the growth rate scaling.

In summary, our numerical simulations provide a
comprehensive understanding of the Tayler instability in spherical geometry,
confirming and extending the results of the global linear analysis.
We successfully reproduce modes known from the WKB approximation and, in the 
regime where resistive effects suppress these, reveal a distinct class 
of large-scale fluctuations which exhibit key similarities with 
the global equatorial eigenmodes identified in the linear analysis.
\section{Application to red giant cores}
\label{s:red_giants}
Our direct numerical simulations uncovered two regimes for the
nonaxisymmetric Tayler instability.
The first is characterized by unstable modes localized at high to
intermediate latitudes, with fast growth rates and small radial
length scales, which are also accessible to the WKB approximation.
The second regime consists of global modes, extending in the equatorial
region and exhibiting reduced growth rates and
large radial length scales.
This latter regime lies adjacent to the region
of the parameter space $(k_N,k_\eta)$
where no instability is found.
The stability line given by Eq.~(\ref{e:sim_stab_line}), derived from
our numerical simulation results, can be recast as a
cubic polynomial equation:
\begin{equation}
\label{e:poly_stable}
    x^{3} - a_1\mathcal{A}_1\,x - a_0\mathcal{A}_0 = 0 ,
\end{equation}
where $x=B_\phi^{1/4}$. The coefficients are $a_0=0.71$ and $a_1=5.22$,
and the functions $\mathcal{A}_0$ and $\mathcal{A}_1$
are defined by
\begin{equation}
\label{e:A0}
\mathcal{A}_0 = (4\pi\rho)^{3/8}(N\,\eta)^{1/2} \kappa^{-1/4} R^{1/4}
\end{equation}
and
\begin{equation}
\label{e:A1}
\mathcal{A}_1 = (4\pi\rho)^{1/4} \eta^{1/2}  R^{-1/2} .
\end{equation}
All quantities are expressed in CGS units, and
$R$ denotes the radial extent of the radiative region.
The solution to this equation yields the critical
toroidal field strength
for instability onset, $B_{\text{crit}}$, above which the global modes
become the dominant unstable fluctuations.

An analogous equation can be obtained for the transition line
between the global mode regime and the regime of WKB-type modes (Eq.~\ref{e:sim_glob_line}).
In this case, the coefficients are $a_0=1.12$ and $a_1=9.01$,
and its solution provides the transition toroidal
field strength $B_{\text{trans}}$.
For a given star, the knowledge of the size of its radiative region $R$,
together with radial profiles of the density $\rho$, thermal
Brunt-V\"{a}is\"{a}l\"{a} frequency $N$, thermal diffusivity $\kappa$,
and magnetic diffusivity $\eta$ therein,
allows us to predict these two threshold field strengths.

\begin{figure}[t]
\centering
\includegraphics[width=\hsize]{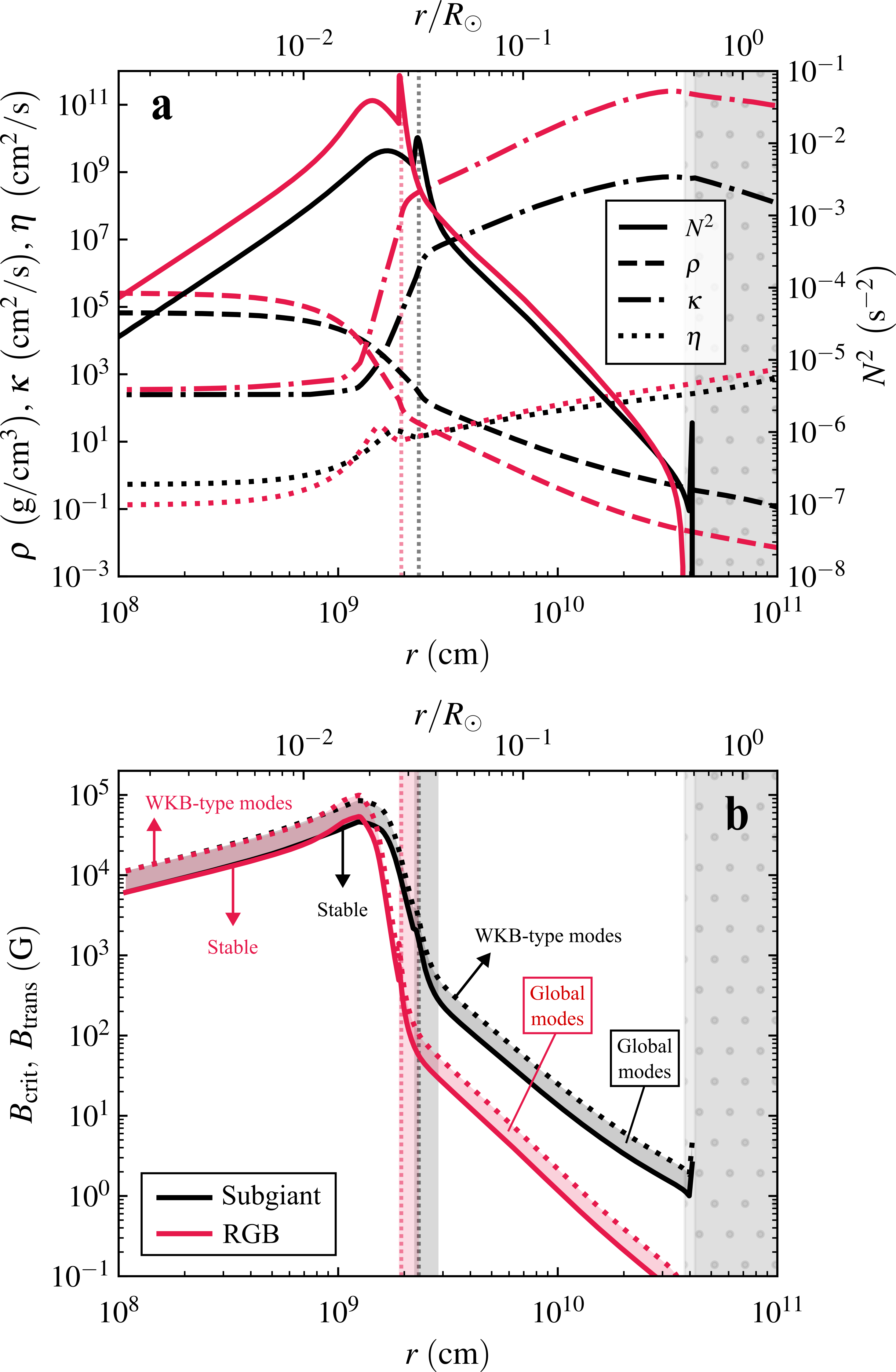}
\caption{Radial profiles of (a)~key stellar quantities
and (b)~threshold toroidal field strengths
for the $1.1\,M_\odot$ stellar model at subgiant (black lines)
and red giant branch (red lines) stages.
The profiles span the radiative cores and the lower parts
of the convective envelopes (vertical gray dot-hatched regions).
Vertical dotted lines mark the approximate locations
of the hydrogen-burning shell.
The upper horizontal axis shows the radius $r$
in units of the solar radius $R_\odot$; the surface
radii are $3.2 R_\odot$ (subgiant) and $16.1\,R_\odot$ (red giant).
Panel~(a) displays the thermal contribution to the squared Brunt-V\"{a}is\"{a}l\"{a}
frequency ($N^2$; right axis), along with the density $\rho$, thermal diffusivity $\kappa$,
and magnetic diffusivity $\eta$ (left axis).
Panel~(b) shows the critical toroidal field strength for instability
onset ($B_\text{crit}$, solid lines) and the transition field strength
($B_\text{trans}$, dotted lines) marking the boundary between the global
and WKB-type instability regimes.
The shaded regions between the curves indicate the
range of field strengths where global modes are expected.
The vertical shaded region near the hydrogen-burning shell
of each star highlights the region where the chemical contribution to the
Brunt-V\"{a}is\"{a}l\"{a} frequency, $N_\mu$, becomes significant
compared to the thermal one, i.e., where $|N_\mu|>N/2$.
}
\label{f:Bcrit}
\end{figure}
To estimate $B_\text{crit}$ and $B_\text{trans}$,
we computed stellar evolution models
of two stars with initial masses of $1.1$ and $1.5~M_\odot$
(where $M_\odot$ is the mass of the Sun). These models span
the mass range of subgiants and red giants for which internal
magnetic fields have been detected (Sect.~\ref{s:intro}).
We evolved both stars from the zero age main sequence up to the red giant branch using
the Catania version of the Garching Stellar Evolution Code \citep{Weiss08,Bonanno02},
reaching final ages of $10.53$ and $3.34\text{ Gyr}$ for the $1.1$ and $1.5~M_\odot$
stars, respectively.
We assumed an initial hydrogen and helium mass fraction of $X=0.70$ and $Y=0.27$, respectively.
No heavy-element diffusion was included and we chose a mixing-length
parameter $\alpha_\text{MLT} = 1.65$, obtained from solar calibration.
For each stellar mass, we selected two models:
one in the subgiant (SG) phase
and one at the base of the red giant branch (RGB).
Since we found no significant difference in our estimates of
$B_\text{crit}$ and $B_\text{trans}$ for the two stellar masses,
we discuss only the results for the $1.1\,M_\odot$ star in the following.
For this star, the SG (RGB) model has an age of $10.127$ ($10.489$) Gyr,
a radius of $R_\ast/R_\odot=3.2$ ($16.1$),
and a luminosity of $L_\ast/L_\odot=4.6$ ($70.1$), where $R_\odot$ and $L_\odot$
are the radius and luminosity of the present-day Sun, respectively.

We calculated the thermal diffusivity
$\kappa$ as \citep[e.g.,][]{Cassisi07}
\begin{equation}
    \kappa=\frac{16\,\sigma_\text{B}\,T^3}{3 k \rho^2\,c_p},
\end{equation}
where $\sigma_\text{B}$ is the Stefan-Boltzmann constant,
$c_p$ is the specific heat capacity at constant pressure,
and $k$ is the total opacity. The total opacity is defined
by $k^{-1}=k_\text{rad}^{-1}+k_\text{cond}^{-1}$, with $k_\text{rad}$ and
$k_\text{cond}$ being the radiative and
conductive opacities, respectively, as computed in GARSTEC \citep[see][]{Weiss08}.
The magnetic diffusivity $\eta$ is computed
using standard expressions detailed in Appendix~A, which
account for the collisional contribution from Spitzer conductivity
in the outer, nondegenerate regions of the core, and
for electron degeneracy effects in the inner layers.

Figure~\ref{f:Bcrit}a presents the radial profiles of
the thermal contribution to the squared Brunt-V\"{a}is\"{a}l\"{a} frequency ($N^2$),
density ($\rho$), thermal diffusivity ($\kappa$),
and magnetic diffusivity ($\eta$)
within the radiative cores of the SG (black lines) and
RGB (red lines) stars. These profiles extend
up to the lower part of the convective envelopes ($N^2<0$, indicated by the
gray shaded regions with dotted hatching).
For each star, hydrogen burning takes place within a narrow shell,
the hydrogen-burning shell (HBS). Its approximate location is
indicated by the vertical dotted line, which coincides with
the luminosity jump and aligns with the peak in $N^2$ (solid line).
The regions of the core below the HBS are composed of inert helium.
Here, $N^2$ increases by about an order
of magnitude from the SG to the RGB stage, primarily due to
the rise in density (dashed lines) driven
by the core gravitational contraction.
In these inner regions, the density reaches up to
$10^5\,\text{g}/\text{cm}^3$, and heat transport is dominated
by electron conduction.
In contrast, in the core regions above the HBS,
heat is primarily transported by radiation.
The steep density drop across the HBS leads to an increase
in thermal diffusivity $\kappa$ (dash-dot lines)
by more than 4 orders of magnitude in this region.
For both stellar stages, the magnetic diffusivity $\eta$
remains low -- between 0.1 and $1\,\text{cm}^2/\text{s}$
-- in the inner degenerate core regions due to the high conductivity of
the electron gas, and rises to $10^2-10^3\,\text{cm}^2/\text{s}$
in the outer, nondegenerate parts of the core (dotted lines).

Figure~\ref{f:Bcrit}b displays $B_\text{crit}$ (solid lines)
and $B_\text{trans}$ (dotted lines) as a function of radius
for the SG and RGB stars.
Toroidal fields with strengths $B_\phi < B_\text{crit}$
are stable. In the degenerate core regions below
the HBS, the profiles of $B_\text{crit}$ are nearly identical for
the two stars. They increase monotonically from $6\,\text{kG}$
in the innermost parts to about $50\,\text{kG}$ as they approach the HBS.
Just below the HBS, $B_\text{crit}$ drops sharply
and continues to decline in the outer core regions, reaching values
of a few $\text{G}$ or less near the base of the convection zone.
We caution that our analysis does not strictly apply in a narrow region
in the vicinity of the HBS, where chemical stratification -- neglected here --
becomes comparable to or stronger than thermal stratification
(see vertical gray and red shaded regions in Fig.~\ref{f:Bcrit}b).

Global Tayler modes develop within a limited range of field strengths,
specifically where $B_\text{crit} < B_\phi \leq B_\text{trans}$
(gray and red shaded areas in Fig.~\ref{f:Bcrit}b).
Our numerical simulations and linear analysis confirm
the relation of BU12 for the growth rate of the global modes,
\begin{equation}
\label{e:BU12_gam}
    \Gamma \propto \epsilon\,\delta^{-2}
\end{equation}
or, in CGS units,
\begin{equation}
  \sigma \propto \frac{B_\phi^2\,\kappa}{4\pi\rho R^4 N^2} .
\end{equation}
Here, we adopted the definitions from Sec.~\ref{s:lin_num}, with
$R$ denoting the characteristic radial extent of the unstable region
within the core where global modes operate.
The proportionality factor in (\ref{e:BU12_gam}) is determined
from the numerical simulations in subset~1 (see Fig.~\ref{f:growth_rate}
and the corresponding discussion).
The $e$-folding time of these modes, $\tau=\sigma^{-1}$,
varies significantly across the radiative core and
depends sensitively on $R$, which itself reflects the evolutionary stage.
For instance, assuming $R=10^9\,\text{cm}$, $\tau$ remains several orders of magnitude
higher than the core evolution timescale -- hundreds of Myr for subgiants
and tens of Myr on the RGB --
in regions below the HBS, rendering global modes dynamically irrelevant there.
However, in the outer core layers where stratification is weaker, $\tau$
becomes comparable to evolutionary timescales, indicating that global
modes may have dynamical impact.

The radial profiles of $B_\text{trans}$ for the two stellar stages
broadly follow those of the critical field strengths
discussed above, but with systematically higher amplitudes
(Fig.~\ref{f:Bcrit}b, dotted lines).
Below the HBS, $B_\text{trans}$ ranges
between $10$ and $100$~kG for both stellar stages.
Above the HBS, $B_\text{trans}$ decreases monotonically
with radius to values of the order of $1\,\text{G}$ or below
near the base of the convection zone.
Dynamical impact on the core is expected in the regime
of WKB-type modes, corresponding to toroidal field strengths exceeding $B_\text{trans}$.
These modes grow at Alfv\'enic rates, corresponding to core Afv\'en travel
times of $10^2-10^3$ years for field strengths near $B_\text{trans}$.
\section{Summary and conclusions}
\label{s:conclusions}
We first investigated the current-driven
nonaxisymmetric ($m=1$) Tayler instability
in a spherical domain using a linear stability analysis that is
global in radius.
This analysis accounts for stable thermal stratification and thermal diffusion,
and extends the work of \citet{Bonanno12a}
by including the latitudinal structure of the background toroidal field.
The background toroidal field is in magnetohydrostatic equilibrium,
with gravitational, pressure, and Lorentz forces in balance, as
expected in radiative stellar interiors.
Our global approach captures the full radial structure
of the unstable $m=1$ perturbations,
a feature inaccessible to fully local analyses
\citep{Acheson78,Spruit99,Skoutnev24},
and demonstrates the stabilizing influence of gravity on
these perturbations.

Next, we used the same background toroidal field to study
the Tayler instability through 3D direct numerical simulations
in a nearly full-sphere geometry,
tracking its evolution from the linear to the nonlinear phase.
Unlike previous studies in a similar geometry \citep{Guerrero19,Monteiro23},
our simulations systematically explore
the effect of thermal stratification on the instability,
including the influence of finite viscosity and magnetic resistivity.
The numerical results revealed two distinct instability regimes,
each characterized by a different class of nonaxisymmetric fluctuations.
The simulations allowed us to derive empirical laws for both the critical
toroidal field strength for instability onset and the
transition field strength separating the two instability regimes, expressed
in terms of fundamental stellar fluid properties.

In the highly supercritical regime where viscous and resistive effects
are negligible, both the linear analysis and the numerical
simulations confirm previous results from local analyses
in the WKB approximation.
The most unstable modes in this regime are confined between mid to high latitudes,
forming a wedge-shaped region around the vertical symmetry axis of
the background toroidal field.
As the stratification parameter $\delta$ (defined as the ratio
of the Brunt-V\"ais\"al\"a frequency $N$ to the Alfv\'en frequency
of the background toroidal field $\omega_{\text{A}0}$)
increases, the radial wavenumber of these modes increases, while their
growth rate remains on the order of $\omega_{\text{A}0}$,
as expected.

In the equatorial region, our linear analysis unveils
a new class of unstable modes, inaccessible to a fully local approach,
characterized by large radial length scales and reduced growth rates.
The numerical simulations demonstrate that these global modes
dominate the dynamics when the system approaches the stability line, specifically
when viscous and resistive effects start to dampen the
high-latitude WKB-type modes.
Both our linear analysis and, for the first time, our numerical simulations
confirm that the dimensionless growth rate of the global equatorial modes
follows a scaling of
$\Gamma \sim \delta^{-2}$, or equivalently,
in dimensional quantities, $\sigma \sim
\omega_{\text{A}0}^3/N^2$, as anticipated by BU12.

Our results provide a general framework for evaluating the linear stability
of toroidal magnetic fields within radiative stellar interiors.
From our numerical simulations, we established
two straightforward algebraic equations (Eqs.~\ref{e:poly_stable}-\ref{e:A1})
relating the critical toroidal field strength
for instability onset $B_\text{crit}$, and the transition field strength $B_\text{trans}$, which
separates the unstable regimes where global and WKB-type modes dominate,
to internal fluid properties readily available from stellar evolution models.

We computed stellar evolution models for a $1.1$ and $1.5\,M_\odot$ stars
and selected two evolutionary stages for each star: the early subgiant phase
and the base of the RGB.
These models provided radial profiles of the
Brunt-V\"ais\"al\"a frequency $N$,
density $\rho$, and thermal diffusivity $\kappa$ in the radiative core.
We also estimated the magnetic diffusivity $\eta$ in both
the outer nondegenerate and inner degenerate regions of the core.

For both stellar masses, the degenerate core regions below the HBS show
$B_\text{crit}$ ranging from $6$ to $50\,\text{kG}$ and $B_\text{trans}$ from $10$ to $100\,\text{kG}$,
with little variation between the two evolutionary stages.
Both threshold field strengths drop sharply -- by over two orders of magnitude --
across the HBS and continue to decline outward
to values below $1\,\text{G}$. This trend suggests that the outer regions
of subgiant and red giant cores are likely to host unstable toroidal fields.
We recall that these results apply in the limit of slow
rotation, i.e., when the rotation rate $\Omega$ is much smaller
than $\omega_{\text{A}0}$ locally, as the stabilizing influence
of the Coriolis force on the instability \citep[e.g.,][]{Spruit99}
was not taken into account.

Recent asteroseismic studies suggest that axisymmetric
toroidal fields may dominate, though not overwhelmingly,
over the detected radial magnetic fields
in the cores of subgiants and red giants \citep{Li22,Deheuvels23,Li23}.
However, detecting such toroidal fields remains challenging,
as asteroseismic inversion kernels exhibit limited sensitivity
to them, at least for the background configurations
explored to date \citep{Bhattacharya24}.
Broader investigations of alternative field configurations are
needed to clarify the detectability of toroidal fields using
mixed oscillation modes. Despite these limitations,
the threshold toroidal field strengths
derived here offer valuable constraints for future asteroseismic
studies, particularly in core regions where rotation plays
a minor role.

We emphasize that our results apply to radiative interiors across a wide
range of stellar masses and evolutionary stages.
For the core of the present-day Sun,
for instance, the Brunt-V\"{a}is\"{a}l\"{a} frequency
is of approximately $N = 10^{-3}\,\text{s}^{-1}$, with thermal and magnetic diffusivities
of $\kappa\sim 10^8\,\text{cm}^2/\text{s}$
and $\eta\sim 10^3\,\text{cm}^2/\text{s}$, respectively.
Using our scaling relations, we estimate a critical field strength
of $B_\text{crit}\sim 10\,\text{G}$ and a transitional strength
of $B_\text{trans}\sim 10^2\,\text{G}$.

While a detailed characterization of the nonlinear solutions
is left for future work, our simulations already provide insight into
the resulting magnetic field geometries. In the global mode regime,
the field evolves toward a simple, large-scale configuration
dominated by the axisymmetric toroidal component.
This is because the nonaxisymmetric field fluctuations grow so slowly
that the background toroidal field decays resistively before reaching
the nonlinear stage, and the axisymmetric poloidal field generated
by the instability remains negligible.
In contrast, in the regime of WKB-type modes, the instability generally produces
turbulent field solutions with mixed poloidal-toroidal character,
owing to flow and field fluctuations growing at a faster Alfv\'en rate.
In our most strongly stratified and supercritical run (TS12; see Table~\ref{t:sim_runs}),
the ratio of the RMS axisymmetric poloidal to toroidal field reaches
about $10^{-2}$ in the nonlinear phase.

In this study, we have not addressed the effect of rotation
on the instability, leaving this important aspect for future work.
While local perturbation analyses
indicate that the Coriolis force generally reduces the growth rate
of the Tayler instability compared to the nonrotating
case \citep{Pitts85,Spruit99,Skoutnev24}, a comprehensive understanding
of the instability and the resulting turbulent transport
under the combined influences of rotation, differential rotation,
and both chemical and thermal stratification -- all relevant
to stellar interiors -- is lacking.
Nonetheless, progress has recently begun in this direction.
Local linear analysis incorporated
the effects of rotation and of both thermal and compositional stratification
in spherical geometry \citep{Skoutnev25},
while current numerical efforts are devoted to test the scaling predictions
for turbulent transport derived from order-of-magnitude estimates
of the instability saturation \citep{Petitdemange23,Barrere23,Barrere25}.
Extending numerical simulations to strongly stratified regimes, as initiated
in this work, together with global linear stability analyses like the one presented here,
is crucial for understanding the saturation mechanism of the Tayler instability
and accurately quantifying the associated turbulent transport.
\begin{acknowledgements}
      The authors are grateful to the anonymous reviewer for their insightful and constructive comments, which helped improve this manuscript. This research was funded by the Deut\-sche For\-schungs\-ge\-mein\-schaft (DFG, German Research Foundation) - AR~355/13-1, within the framework of ``Stability and generation of magnetic fields in red giants: towards a unified picture''. DGM acknowledges support from the European Union - NextGenerationEU within the framework of the Italian National Recovery and Resilience Plan (NRRP), Mission 4: Education and Research, Component M4C2, Investment 1.2, through the ``Bando Young Researcher 2024'' Project Number~SOE2024\textunderscore0000097 ``Elucidating the Transport of Angular momentum by MAgnetic fields in red GIAnts (ETAMAGIA)''. AB acknowledges support from the European Union – NextGenerationEU RRF M4C2 1.1  n: 2022HY2NSX ``CHRONOS: adjusting the clock(s) to unveil the CHRONO-chemo-dynamical Structure of the Galaxy''. AB and GL acknowledge support from the research grant ``Unveiling the magnetic side of the Stars'' funded under the INAF national call for Fundamental Research 2023.
\end{acknowledgements}
\bibliographystyle{aa}
\bibliography{biblio_TI}
\begin{appendix}
\section{Microscopic magnetic diffusivity}
\label{app:eta}
The magnetic diffusivity in red giant cores is computed using standard expressions detailed below.
We define the total magnetic diffusivity, $\eta$, as
\begin{equation}
    \eta = \frac{c^2}{4\pi(\sigma_\text{coll}+\sigma_e)},
\end{equation}
where $c$ is the speed of light, and $\sigma_\text{coll}$ and $\sigma_e$
denote the contributions to the electrical conductivity from collisional processes and electron degeneracy, respectively.
For $\sigma_\text{coll}$, which dominates over $\sigma_e$ in the core regions
above the hydrogen-burning shell, we adopt the classical
expression for the electrical conductivity of a nondegenerate, fully ionized, ideal gas of \cite{Spitzer62}
\begin{equation}
\label{e:A_sigma_coll}
    \sigma_\text{coll} = \gamma_\text{E} \frac{2\,(2 k_\text{B} T)^{3/2}}{\pi^{3/2}m_e^{1/2} Z e^2 \ln\Lambda},
\end{equation}
where $k_\text{B}$ is the Boltzmann constant, $T$ is temperature, $m_e$ is the electron mass, $Z$ is the atomic number, $e$ is the electron charge, $\ln\Lambda$ is the Coulomb logarithm, and $\gamma_\text{E}$ is the correction for electron-electron scattering which we set to 0.582 for a pure hydrogen plasma \citep{Wendell87}.

Below the hydrogen-burning shell in red giant cores, electron degeneracy becomes significant and transport processes are governed by electron conduction \citep{Garaud15}. The electrical conductivity is dominated by
the degenerate electron contribution $\sigma_e$ here, for which we adopt the expressions
derived by \cite{Hubbard66} for a strongly degenerate electron gas.
The corresponding thermal conductivity $K$ is given by
\begin{equation}
\label{e:A_K_e}
    K = \frac{(2\pi \hbar)^3 k_\text{B}^2}{16 m_e^2 e^4 A m_H}\, G_\gamma(\kappa_\text{F}) \rho T
\end{equation}
where $\hbar$ is the reduced Planck constant, $\rho$ is density, $A$ is the average atomic number of the nuclei,
$m_H$ is the mass of the hydrogen atom, and $\kappa_\text{F}=(9\pi Z/4)^{1/3}$ is the dimensionless Fermi wavenumber.
The function $G_\gamma$ is
\begin{equation}
G_\gamma (x) = \frac{1}{\ln(1+4x^2/3\gamma)},
\end{equation}
where $\gamma$ is the dimensionless parameter
\begin{equation}
\label{e:A_gam}
    \gamma = \frac{Z^2 e^2}{k_\text{B} T}\left( \frac{4\pi\rho}{3 A\, m_H}\right)^{1/3} .
\end{equation}
Since the degenerate red giant core regions
are mostly composed of helium,
we use $A=4$ and $Z=2$ in Eqs.~(\ref{e:A_K_e}) and (\ref{e:A_gam}).
In the limit of high degeneracy, the Wiedemann-Franz law holds,
\begin{equation}
\label{e:A_WF}
K = \frac{1}{3}\pi^2 k_\text{B}^2 \sigma_e T /e^2 ,
\end{equation}
which we use to determine $\sigma_e$ from the thermal conductivity $K$.
In the region of weak electron degeneracy around the hydrogen-burning shell, where neither the expression
for $\sigma_e$ nor that for $\sigma_\text{coll}$ is strictly applicable,
we estimate the magnetic diffusivity $\eta$ using
a cubic spline interpolation between the degenerate contribution
from the deeper core and the collisional contribution from the outer core.
\end{appendix}

\end{document}